\documentclass[letterpaper,pdftex,11pt]{article}
\usepackage{interspeech2010,amssymb,amsmath,epsfig}
\usepackage[safe]{tipa}
\usepackage{tipx,nicefrac,url}
\usepackage{subfigure}

\def\reg{{\rm\ooalign{\hfil
     \raise.07ex\hbox{\scriptsize R}\hfil\crcr\mathhexbox20D}}}

\title{Open-loop multi-channel inversion of room impulse response}

\usepackage{tabularx}
\makeatletter
\def\hlinewd#1{%
\noalign{\ifnum0=`}\fi\hrule \@height #1 %
\futurelet\reserved@a\@xhline}
\def\name#1{\gdef\@name{#1\\}}
\makeatother
\name{{\em Bowon Lee, Camille Goudeseune, Mark A. Hasegawa-Johnson}}
\address{University of Illinois at Urbana-Champaign \\
{\small \tt \{bowonlee,cog,jhasegaw\}@illinois.edu}}

\begin{document}
\maketitle
{
\custompt
\begin{abstract}

This paper considers methods for audio display in a CAVE-type virtual
reality theater, a 3 m cube with displays covering all six rigid faces.
Headphones are possible since the user's headgear continuously
measures ear positions, but loudspeakers are preferable since they
enhance the sense of total immersion.  The proposed solution consists
of open-loop acoustic point control. The transfer function, a matrix of
room frequency responses from the loudspeakers to the ears of the user,
is inverted using multi-channel inversion methods, to create exactly
the desired sound field at the user's ears.  The inverse transfer
function is constructed from impulse responses simulated by the image
source method. This technique is validated by measuring a $2 \times 2$
matrix transfer function, simulating a transfer function with the same
geometry, and filtering the measured transfer function through the
inverse of the simulation. Since accuracy of the image source method
decreases with time, inversion performance is improved by windowing the
simulated response prior to inversion. Parameters of the simulation
and inversion are adjusted to minimize residual reverberant energy;
the best-case dereverberation ratio is 10 dB.

\end{abstract}

\section{Introduction} 
\label{sec:intro}

The task of interest in this paper is free-field audio display for a virtual reality environment~\cite{Cruz-Niera1993}.  The virtual reality testbed for these experiments is a 3 m cube called ALICE (A Laboratory for Interactive Cognitive Experiments), located at the University of Illinois.  Images are projected from outside onto all faces of the cube.  Users are untethered: no wires connect equipment they wear to the outside world. In order to accurately convey images to the user, the positions of up to 20 user points (e.g., head, ears, hands, and feet) are precisely tracked using a magnetic tracking system (calibrated mean accuracy of 8 cm and 1 degree, 120 samples updated per second).  The goal of virtual reality in ALICE is total immersion: users must be able to ``suspend disbelief'' and convince themselves that they are physically present in the virtual environment portrayed to them.

The goal of most previous virtual reality audio experiments is to correctly portray the position of a sound source.  Position accuracy is usually achieved by filtering the audio signal through head-related transfer functions and then playing it over headphones.  The disadvantage of headphone audio is that it sounds like it is coming from the headphones.  The impression of total immersion is lost if the audio display is part of the user's headgear rather than part of the environment.

A measure of headphone-free realism is possible by simply playing the desired audio from the most appropriate speaker in a large speaker array.  For the ALICE environment, an array of eight transparent loudspeakers has been prototyped.  These loudspeakers consist of millimeter-thick sheet glass suspended into the cube, connected to compression drivers located outside the walls of the cube.  The transparent speakers provide reasonable audio display with good localization for distant objects (outside the cube wall), and most important, the transparent loudspeakers do not obstruct or distort the video display.

Moving the virtual audio source inside the room is much more difficult.  The positions of the user's ears are known precisely.  If the room impulse response were known, then the known room impulse response could be inverted using well-studied multi-channel inversion methods~\cite{Miyoshi1988,Kirkeby1998},
thus creating exactly the desired sound field at the two ears of the user. Unfortunately the room impulse response is not known.  The user is free to put his or her head anywhere in the room; it is impossible to measure the room impulse response from every speaker location to every possible location of the user's ears.

Two solutions to this problem are possible.  First, an estimate of the room impulse response can be adaptively updated using microphones placed on the user's headgear, by means of a number of adaptive signal processing methods. This paper analyzes a second solution to the problem of headphone-free virtual reality audio display.  The proposed solution consists of open-loop acoustic point control, using a simulation of room impulse response based only on knowledge of the room geometry, architectural materials, and user location.

The image source method of room response simulation was originally proposed for open-loop dereverberation experiments similar to the one proposed here~\cite{Allen1979}. Its performance was never quantitatively reported in the literature, since multi-channel inversion methods for non-minimum phase impulse responses were not well understood at that time~\cite{Neely1979}. Other methods of simulating room impulse response are almost always evaluated by purely qualitative means like acoustic perceptual studies and visual comparisons of impulse responses~\cite{Omoto2002}.

This paper proposes instead to evaluate simulated room impulse responses based on their performance in a regularized dereverberation task.  Dereverberation performance is measured in terms of the decibel ratio of the energy of the room impulse response to that of the dereverberated response.  It is demonstrated that this method can be used to optimize parameters of the model including absorptivity and window taper.

This paper is organized as follows.  Section~\ref{sec:back} describes previous published research in the fields of room impulse response measurement, room impulse response simulation, and room impulse response inversion.  Section~\ref{sec:methods} describes the methods of these tasks in a simulated virtual reality environment, a 2 m plywood cube. Room impulse responses are measured with a starter pistol. An evaluation metric is proposed to quantitatively measure the dereverberation ratio.  Section~\ref{sec:results} discusses the results, demonstrating how the proposed evaluation metric optimizes methods for simulating and inverting the room impulse reponse. Section~\ref{sec:conclusion} reviews conclusions.

\section{Background}
\label{sec:back}

\subsection{Measurement of room impulse response}
\label{sec:back:msrmnt}

An excitation signal is required in order to measure a room impulse response.
A perfect impulse (a Dirac delta function) simplifies the measurement task
(measured response equals the impulse response), but it is not possible to
physically generate a Dirac delta function.
In practice, impulse-like signals or signals with characteristics similar to a
perfect impulse such as flat frequency response are used.

ISO Standard 3382 specifies the following requirements for an excitation signal
for measuring room impulse response~\cite{ISO-3382}.
First, it should be nearly omnidirectional.
Second, its sound pressure level should provide sufficient 
dynamic range to avoid contamination by background noise.
Third, the signal should be repeatable.

Impulse-like signals such as the starter pistol, balloon pop, and electric spark
have been traditionally chosen as excitation signals~\cite{ISO-3382,Schroeder1979}.
These impulses are easily generated and have been used to determine rough
characteristics of a room, such as reverberation time.
However, measurement by these impulse methods has not been widely discussed
in the scientific literature for three reasons. 
First, the frequency response of these impulses is not flat (Fig.~\ref{fig:pist-amb}).
Second, some authors report that it is difficult to get an adequate
SNR because all the impulse's energy is packed into a very 
short duration~\cite{Schroeder1979}.
Third, the signal is not precisely repeatable because it depends significantly on
small variations in charge distribution, balloon shape, etc (Fig.~\ref{fig:pist-shots}).

\begin{figure}
\centerline{\includegraphics[width=5in]{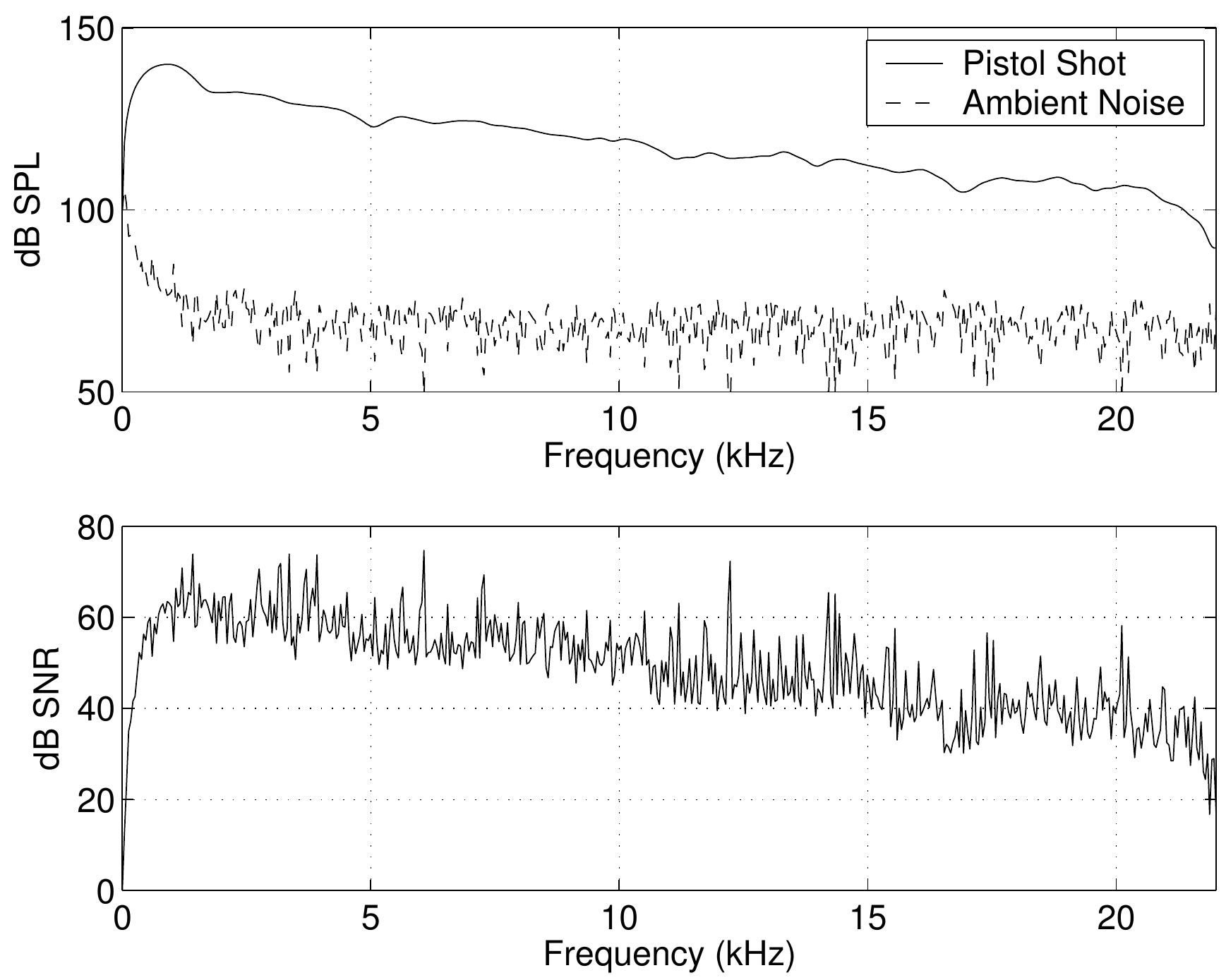}}
\caption{ Magnitude responses of starter pistol and ambient noise.}
\label{fig:pist-amb}
\end{figure}

\begin{figure}
\centerline{\includegraphics[width=4.5in]{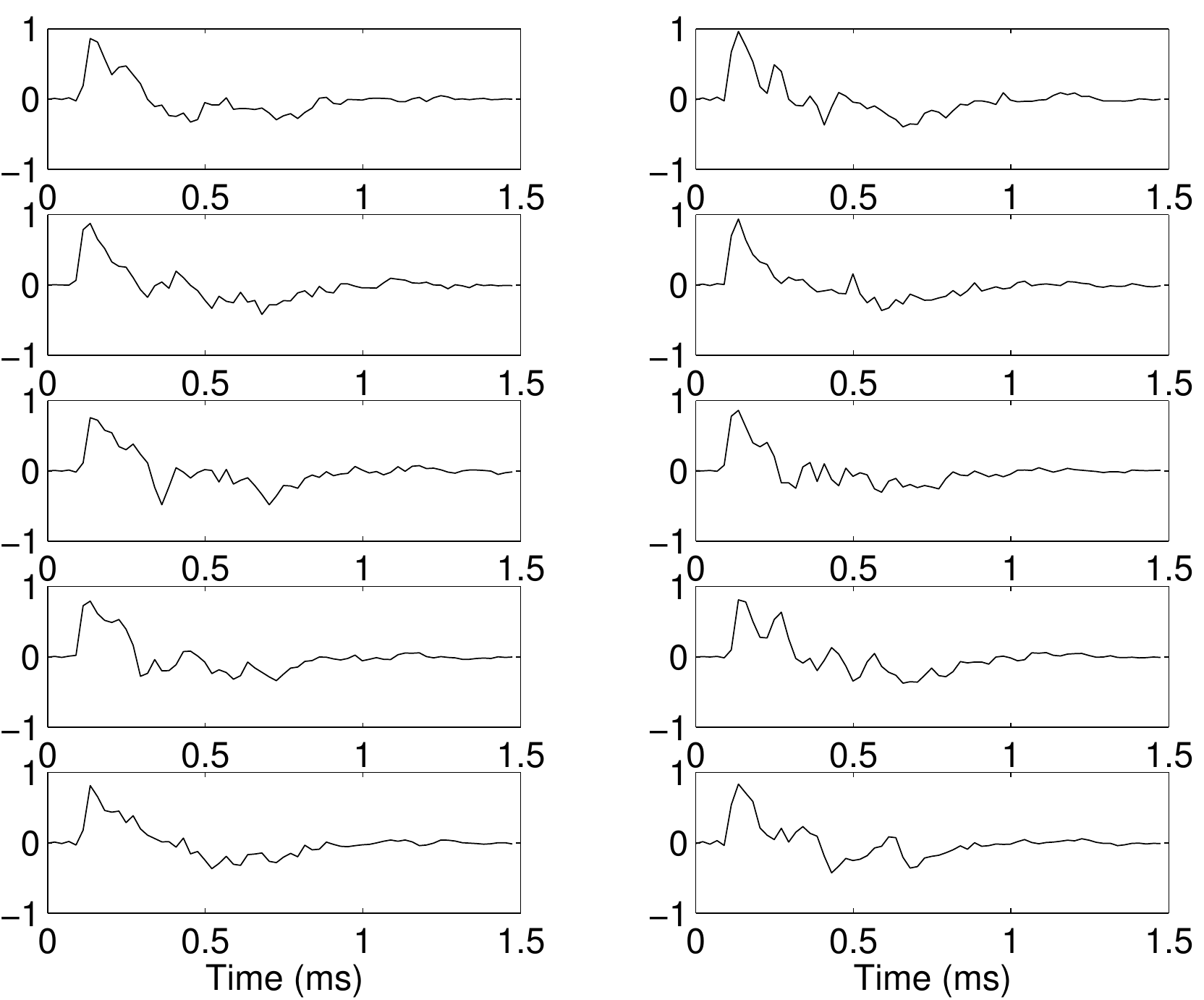}}
\caption{ Waveforms of individual pistol shots.}
\label{fig:pist-shots}
\end{figure}

In 1979 Schroeder suggested an alternative method to measure room impulse response
using Maximum Length Sequences (MLS)~\cite{Schroeder1979}.
The autocorrelation function of an MLS of order $m$ with length $N=2^m-1$ samples
is two-valued, $1$ at time zero and $-1/N$ at times other than zero (modulo $N$).
If $N$ is sufficiently large, then $-1/N$ becomes negligible and
we can assume that the resulting MLS has the same autocorrelation as
pseudo-random noise.
Because MLS signal energy grows with $N$, its SNR can be made
arbitrarily
high without needing high amplitude, which is not the case with impulses.
Since the MLS is stored in a computer, it can be generated repeatedly.
Therefore MLS meets the last two requirements of ISO 3382 better than
impulse-like signals such as starter pistols and electric sparks. 

Computer-generated pseudo-random sequences have discrete values such as $+1$ and $-1$.
Since it is impossible to make perfectly abrupt transitions between these two values,
distortion occurs and the frequency response of the system must therefore be compensated
to acquire accurate impulse responses.
The technique of MLS measurement has also been proven to be vulnerable to the nonlinearity
of measuring equipment, particularly loudspeakers~\cite{Rife1989}.
Nonlinearities produce repeated distortion peaks in the time domain,
which prevent the integrated energy of the impulse response from falling below
$-30$ dB~\cite{Vanderkooy1994,Stan2002}.
A modification of MLS, the inverse repeated sequence
(IRS), reduces the distortion caused by 
nonlinearities~\cite{Dunn1993,Ream1970,Briggs1966}.
Other papers discuss the accuracy of the MLS method
\cite{Simpson1966,Bleakley1995}, 
its computational complexity
\cite{Alrutz1983,Cohn1977,Davies1966a,Davies1966b,Davies1966c},
and its application to a variety of system response measurements
\cite{Rife1992,Vorlander1997,Burkard1990}.

Aoshima proposed the time-stretched pulses technique, based on the time
expansion and compression of an impulsive signal~\cite{Aoshima1981}.
The purpose of the time-stretched pulse signal is to increase the total energy
of the excitation signal while keeping the frequency response flat.

Berkhous proposed a sine sweep as an excitation signal~\cite{Berkhout1980}.
Farina and Ugolotti introduced a logarithmic sine sweep method using
a different deconvolution method~\cite{Farina2002}.
Farina's detailed method accurately derives
an impulse response from the raw measurement by
separating the linear and nonlinear components of the measured impulse response,
where the strength of nonlinear distortion is measured by
observing the harmonic distortion caused by nonlinearity of the system.

Stan {\it et al}. compare four different room impulse response measurement
techniques: pseudo-random noise (MLS and IRS), time-stretched pulses,
and logarithmic sine sweep~\cite{Stan2002}.
Since the randomized phase of psuedo-random sequences makes them immune to
background noise, MLS and IRS techniques are preferred in noisy environments.
However, parameter optimization is required for high SNR because of nonlinear
distortion. Nevertheless, the achieved SNR is
only 60.5 dB with an MLS order of 16 and single measurement.

Time-stretched pulses and sine sweep methods produce a higher SNR than
the pseudo-random noise techniques, but they require a quiet environment.
The SNR of the time-stretched pulses technique is 77 dB
after precise calibration.
The logarithmic sine sweep method has 80.1 dB SNR.
The benefit of the sine sweep is that unlike the previous methods,
it produces a high SNR without any calibration~\cite{Stan2002}.

\subsection{Simulation of room impulse response}

Simulations of room impulse responses fall into two categories:
spatial mesh methods and ray acoustic methods.

Spatial mesh methods numerically solve the constituents
of the acoustic wave equation, namely the equations of motion and continuity
\cite{Rabenstein1999,Schetelig1998}.
In this method, sound pressure and velocity are computed at a finite number
of points, usually mesh points in a cavity. The differential equations in the
continuous domain are computed as difference equations in the discrete domain. 
This method can simulate diffraction effects, which ray acoustic methods cannot.
Unfortunately, to compute an impulse response at a specific location of interest,
the values of sound pressure and velocity must be
computed over the entire mesh, because the solution at a specific point
depends on those of the adjacent points. Spatial mesh methods are thus far more
computationally expensive than ray acoustic methods.

Ray acoustic methods assume that sound rays are emitted from the sound source,
usually as a spherical wave. Ray paths are then traced using either image source
or ray-tracing methods~\cite{Allen1979,Krokstad1968}. 
The ray-tracing method considers a finite number of rays to
be emitted from the sound source.   These ray paths are traced
and their trajectories summed at points of interest.
Although it requires little computation,
ray tracing is appropriate only for a rough estimate, e.g.,
to compute the first few reflections of the room impulse response.

In 1979 Allen and Berkley showed that the impulse response
of a small rectangular room can be computed using a geometric
``image source'' method~\cite{Allen1979}.
Their model creates an ``image space'' without walls,
in which each echo is modeled as the direct sound from
an image source outside the actual walls of the room.
The first six image sources are reflections of the original
source in the six walls of the room.
The next few image sources are created by reflecting the first
six, and so on (Fig.~\ref{fig:illust}).
At each reflection, the amplitude of the source
is scaled by the wall's reflection coefficient.

\begin{figure}
\centerline{\includegraphics[width=4in]{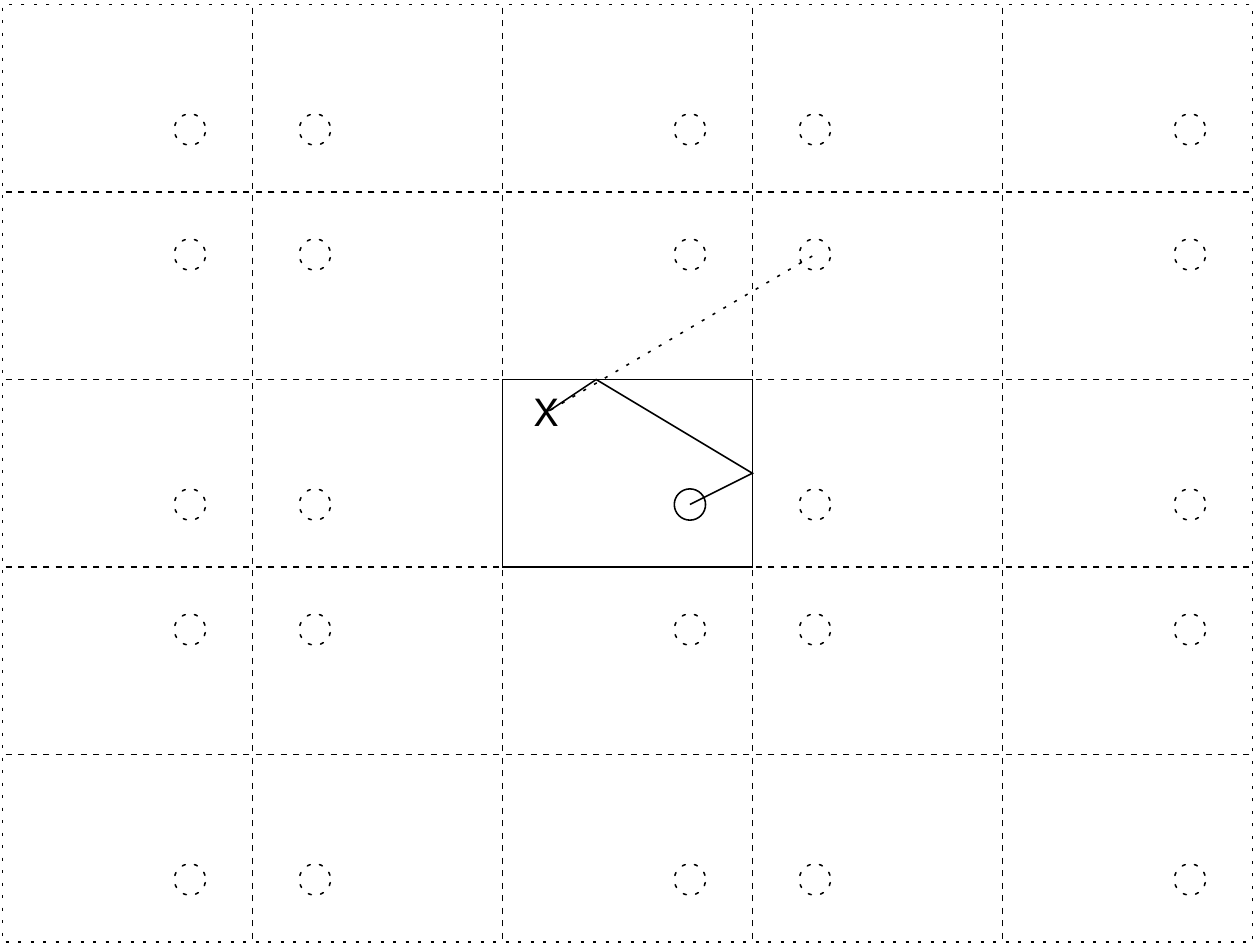}}
\caption{ Two-dimensional illustration of the image source method.; `x'=receiver, solid `o'=source, dashed `o'=image sources.  Solid line=an echo path, dashed line=corresponding image source path.}
\label{fig:illust}
\end{figure}

The image source method requires more computation than ray tracing
because it considers all possible reflected wavefronts.
It can be extended from rectangular cavities to arbitrary polyhedra~\cite{Borish1984}.
In this case, some image sources may not contribute to the total impulse response.
Such image sources are called hidden images.
An algorithm is therefore needed to decide whether a given image source
is hidden or not. Lee and Lee proposed a relatively efficient algorithm for
the image source computation of impulse responses of arbitrarily shaped
rooms~\cite{Lee1988}, but this method is still computationally expensive
relative to the image source method for rectangular rooms.
The image source method is efficient for a rectangular room
because every image source contributes to the total impulse response
(unless there are obstacles in the room), and also, because
the locations of all image sources are analytically pre-computed
due to symmetry.

In a rectangular room, the image sources can be indexed
by integer coordinates $l$, $m$, and $n$, where $(l,\ m,\ n) = (0,\ 0,\ 0)$
corresponds to the direct source, $(1,\ 0,\ 0)$ corresponds
to the first reflection in the positive $x$ direction, and so on.

Given a room of size $(L_x,\ L_y,\ L_z)$ with origin at the center
and a source location $(S_x,\ S_y,\ S_z)$,
the image source location with indices $(l,\ m,\ n)$ is: 
\begin{equation*}
(I_x,\ I_y,\ I_z) = (lL_x+(-1)^lS_x,\ mL_y+(-1)^mS_y,\ nL_z+(-1)^nS_z)
\end{equation*}
Then the distance $d_{lmn}$ from the image source to the receiver
at $(R_x, R_y, R_z)$ is: 
\begin{equation*}
d_{lmn} = \sqrt{(R_x-I_x)^2+(R_y-I_y)^2+(R_z-I_z)^2}
\end{equation*}
The impulse response predicted by the image source method is
\begin{equation}
g(t) = \sum_{l,m,n=-\infty}^{\infty}
       \frac{r^{|l|+|m|+|n|}}{d_{lmn}}\delta(t-\tau_{lmn})
\label{eq:ism}
\end{equation}
where $\tau_{lmn}=d_{lmn}/c$ is the wave propagation time from the
image source at $(l,m,n)$ to the receiver, $c$ is the speed of sound, and
$r$ is the reflection coefficient of the walls.  Eq.~(\ref{eq:ism})
assumes that all surfaces have the same reflection coefficient,
but relaxing this assumption is straightforward and computationally inexpensive.

Three-dimensional audio applications are usually considered in a rectagular cavity,
a room; this paper considers only this special but common case,
to justify use of the otherwise computationally expensive image source method
for simulating the room impulse response.

\subsection{Inversion of room impulse response}

Given the room impulse response, a desired signal can be reproduced
at points of interest if a valid inverse filter
is first created from the impulse response. 
The dereverberation problem thus reduces to constructing such an inverse filter.

Since the purpose of inverting the room impulse response is to
cancel reverberation at multiple points in a room, human ears for example,
the frequency responses and inverse filters are formulated as a matrix of
sequences. Let the term transfer function denote this matrix of frequency
responses. 

Let $G_{ji}(z)$ be the frequency response from the $i^{th}$ loudspeaker
to the $j^{th}$ control point, for $1 \le i \le L$ and $1 \le j \le M$: 
a total of $M \times L$ individual room impulse responses.
The inverse of this transfer fuction is therefore an $L \times M$ matrix. 
The image source method computes the simulated frequency response
$\hat{G}_{ji}(z)$ which approximates $G_{ji}(z)$. 

Let $X_j(z)$ and $\hat{X}_j(z)$ denote the Z-transforms of the
desired and actual control point signals respectively.
The inverse transfer function $H(z)$ has as element $H_{ij}(z)$,
the impulse response from the $j^{th}$ desired control point signal
$X_j(z)$ to the $i^{th}$ loudspeaker signal $V_i(z)$. 
$\hat{X}_j(z)$ is therefore expressed as
\begin{equation*}
\hat{X}_j(z) = \sum_i\hat{G}_{ji}(z)V_i(z)=\sum_{i,k}\hat{G}_{ji}(z)H_{ik}(z)X_k(z)
\end{equation*}
We want to find $H_{ij}(z)$ so that $\hat{X}_j(z)$ is as similar as
possible to $X_j(z)$. Figure~\ref{fig:diagram} shows the diagram of
the room impulse response inversion process.
If $L = M$ and the matrix $\hat{G}(z)$ is minimum phase, then an exact
inverse transfer function is given by $H(e^{j\omega})=\hat{G}(e^{j\omega})^{-1}$.

\begin{figure}
\centerline{\includegraphics[width=4.8in]{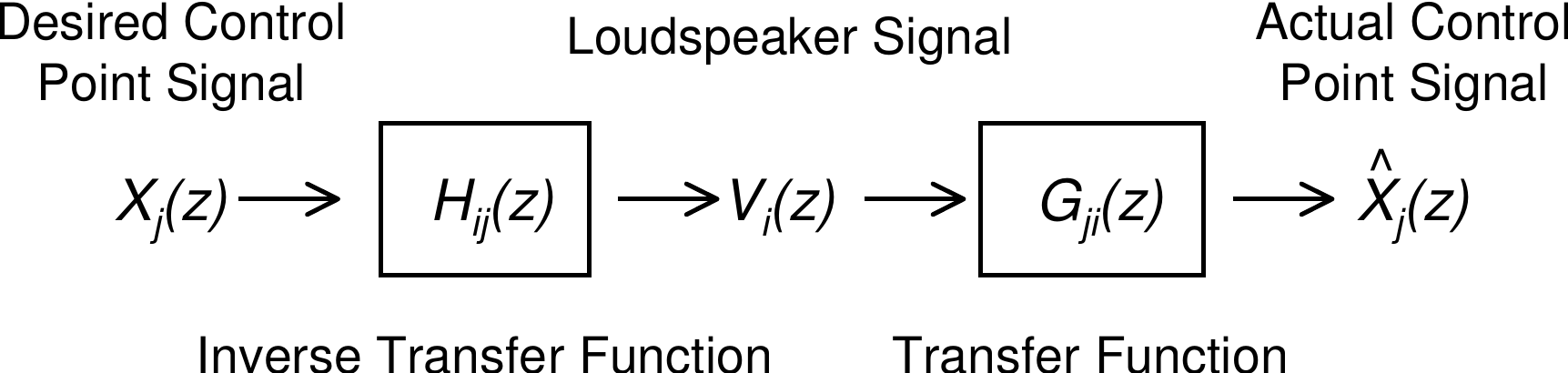}}
\caption{ System diagram.}
\label{fig:diagram}
\end{figure}

If the impulse responses are non-minimum phase, the inverse filter has poles
outside the unit circle.
In this case, we can make the inverse filter either stable
but noncausal (the region of convergence includes the unit circle) or 
causal but unstable (the region of convergence does not include
the unit circle), but not both stable and causal.
Therefore, the exact inverse filter of a square transfer function matrix
is only realizable for minimum phase transfer functions.

Neely and Allen found that the impulse
response of a small room is minimum phase only for reflection
coefficients below approximately 0.37~\cite{Neely1979}.
The impulse response of a small room is rarely
minimum phase, and therefore the stable inverse filter
$\hat{G}(e^{j\omega})^{-1}$ of a square matrix $\hat{G}(e^{j\omega})$
is usually noncausal in practice. 

Miyoshi and Kaneda  showed that the transfer function of a room can be exactly
inverted for the case $L = 2, M = 1$~\cite{Miyoshi1988}.
Nelson {\it et al}.  generalized their result by showing that,
in most circumstances without any extreme symmetry, when $L > M$, the
transfer function can be exactly inverted~\cite{Nelson1992}.
Thus a stable, causal inverse transfer function exists if $G_{ji}(z)$ has more
columns than rows.
Unfortunately no tractable method for finding the causal, stable inverse of
a non-square transfer function in the frequency domain has yet been proposed. 
An equivalent method can be computed in the time domain, but
is computationally expensive.

Recall that a non-minimum phase square transfer function has
a stable but noncausal inverse $H(e^{j\omega})$. A causal, stable
semi-inverse may be constructed by applying a time shift $D$: 
\begin{equation}
\tilde{H}(e^{j\omega}) = e^{-j\omega D}H(e^{j\omega})
\label{eq:moddelay}
\end{equation}
and then truncating $\tilde{h}[n]$ by zeroing the tail at $n < 0$: 
\begin{equation}
\hat{h}[n] = \left\{\begin{array}{ll}\tilde{h}[n],&n\ge0\\
                    0,&n<0\end{array}\right.
\label{eq:trunc}
\end{equation}
This creates a stable and causal approximation
$\hat{h}[n]$ of the exact inverse filter. The time shift $D$ is called
modeling delay.

The inverse transfer function $H$ can be computed by
sampling the spectrum of $\hat{G}$ using an FFT, and inverting
the matrix at each frequency bin.
Sampling the frequency-domain transfer function causes aliasing
in the time domain.
This ``wrap-around effect'' is eliminated by time-shifting $h[n]$
by $e^{-j\omega D}$, which is the same as the modeling delay
described previously.

Merely inverting the sampled FFT matrix $\hat{G}$ yields a poor estimate
of $H$ because of singularities related to the non-minimum phase
character of $\hat{G}$ (zeros of $\hat{G}$ tend to be very close to
the unit circle).  A better estimate can be computed by using
a regularized inversion formula \cite{Kirkeby1998}, in which a small
constant $\beta$ is added to each eigenvalue of $G$ before inversion:
\begin{equation}
H = ({\hat{G}^T}\hat{G} + {\beta}I)^{-1}\hat{G}^T
\label{eq:reginv}
\end{equation}
where $\hat{G}^{T}$ is the Hermitian transpose of $\hat{G}$.

After computing $H(e^{j\omega})$ using Eq.~(\ref{eq:reginv}),
Eqs.~(\ref{eq:moddelay}) and (\ref{eq:trunc}) yield a stable
and causal approximate inverse $\hat{h}[n]$.
The resulting control point signal vector is
\begin{equation*}
\hat{X} = G\hat{H}X
\end{equation*}

Signals produced during regularized inversion are depicted
in Fig.~\ref{fig:proc}. For illustration, a one-dimensional inverse
transfer function is computed from the simulation $\hat{G}$ and
filtered through the simulated transfer function itself
($\hat{X} = \hat{G}\hat{H}X$).
Regularized inversion gives more than $50$ dB of SNR with 
$D=750$ ms and $\beta = 0.05$
(Fig.~\ref{fig:proc}). 

\begin{figure}
\centering
  \subfigure[][]{
    \label{fig:proc:a}
    \includegraphics[width=3.1in]{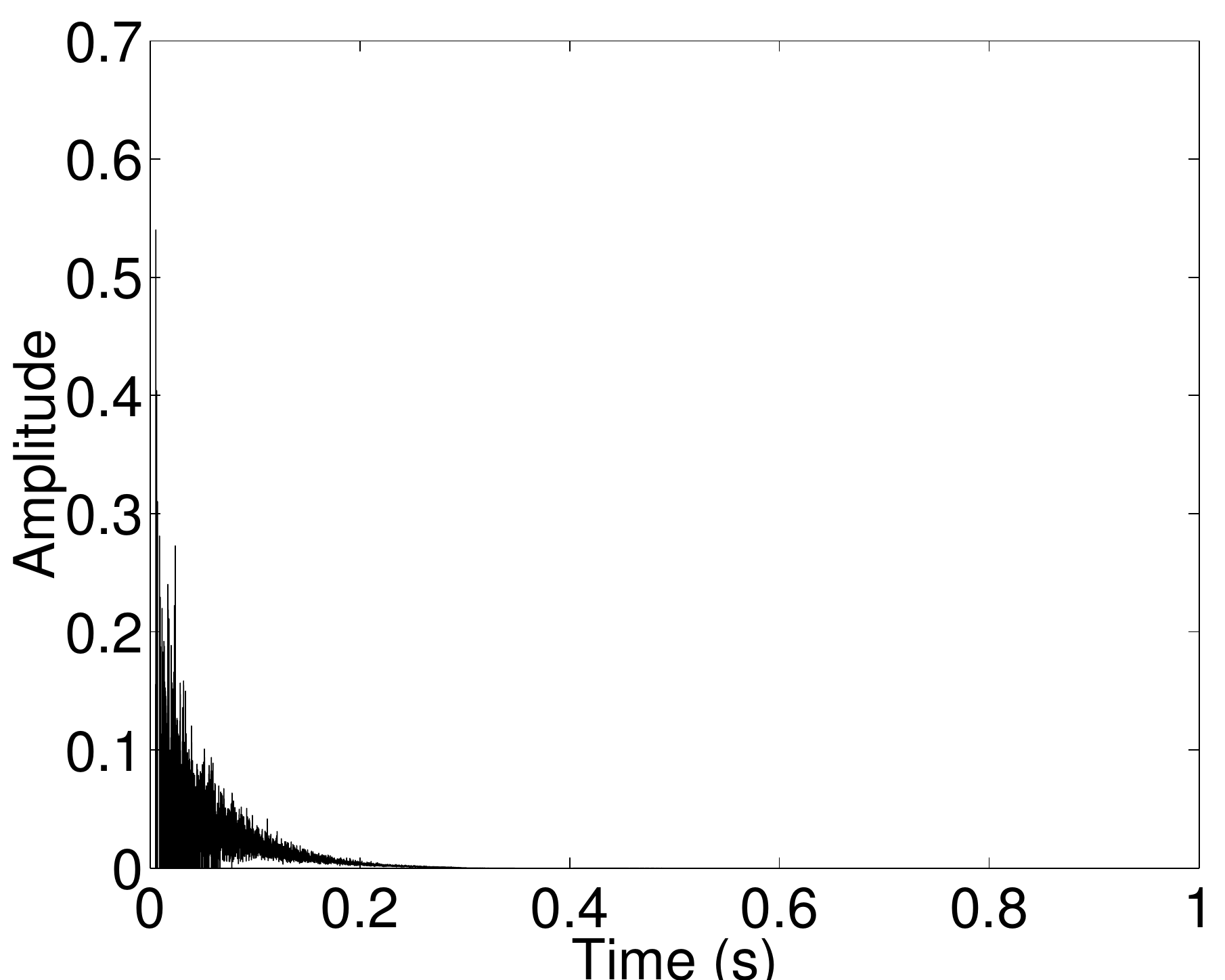}}
  \subfigure[][]{
    \label{fig:proc:b}
    \includegraphics[width=3.1in]{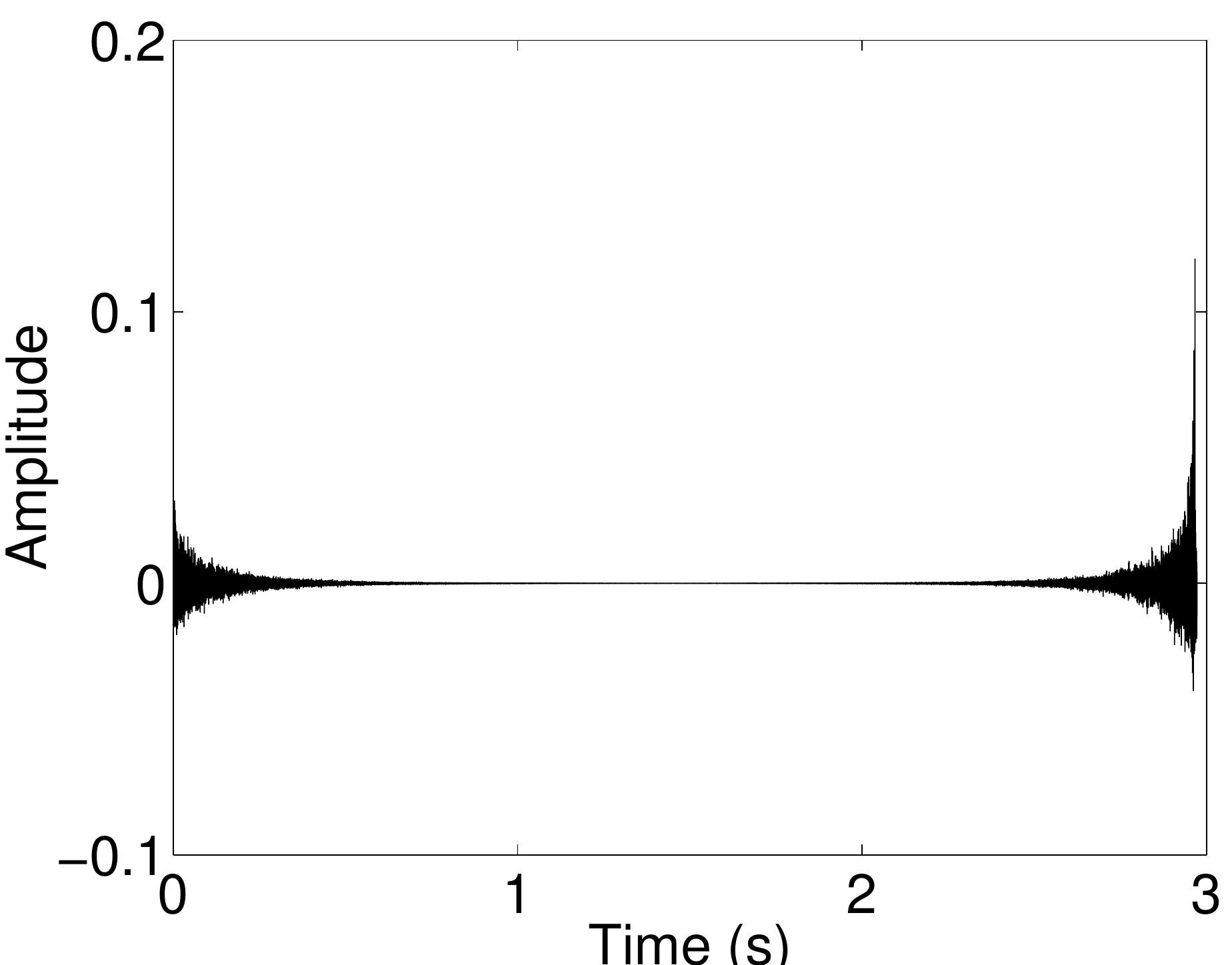}} \\
  \subfigure[][]{
    \label{fig:proc:c}
    \includegraphics[width=3.1in]{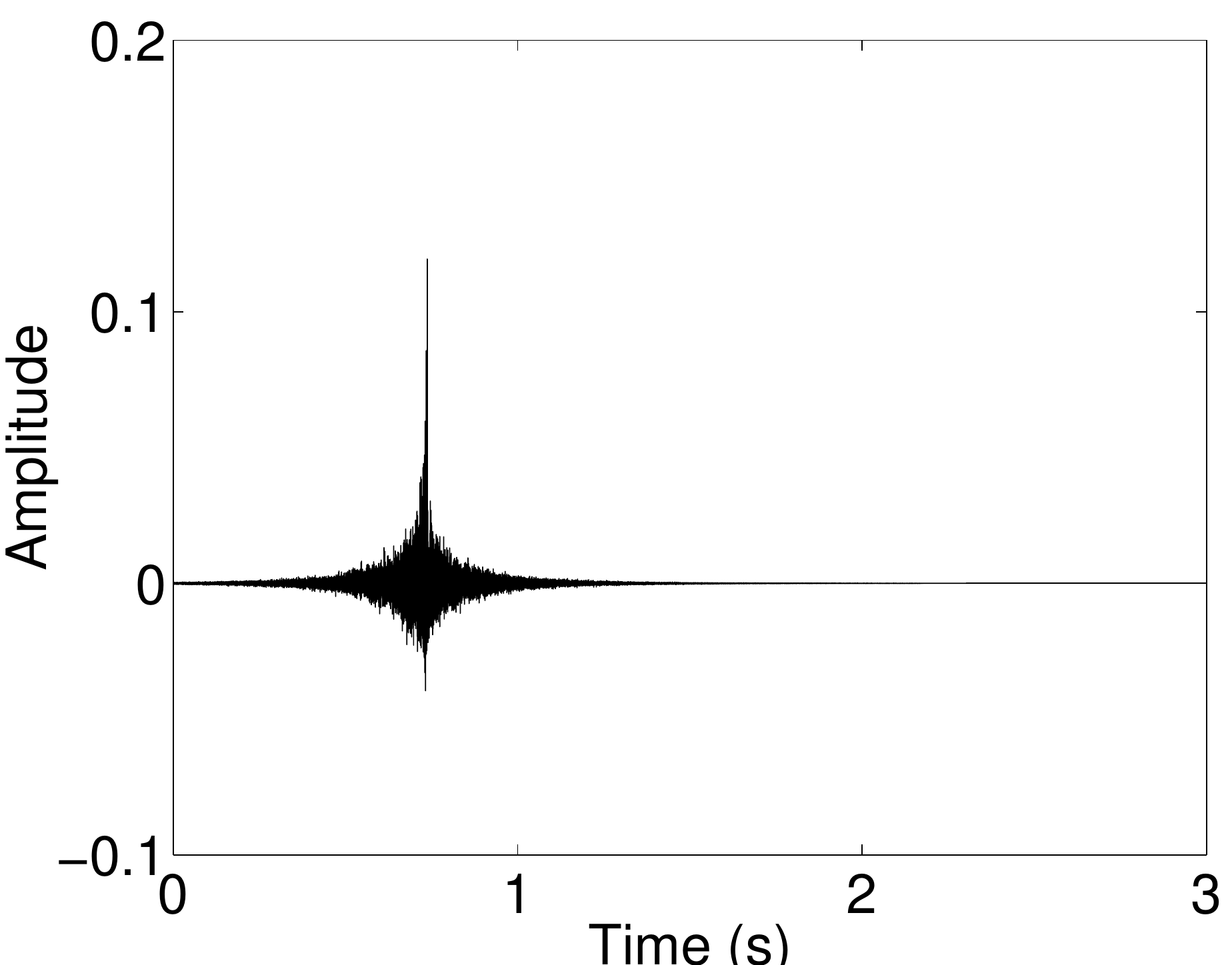}}
  \subfigure[][]{
    \label{fig:proc:d}
    \includegraphics[width=3.1in]{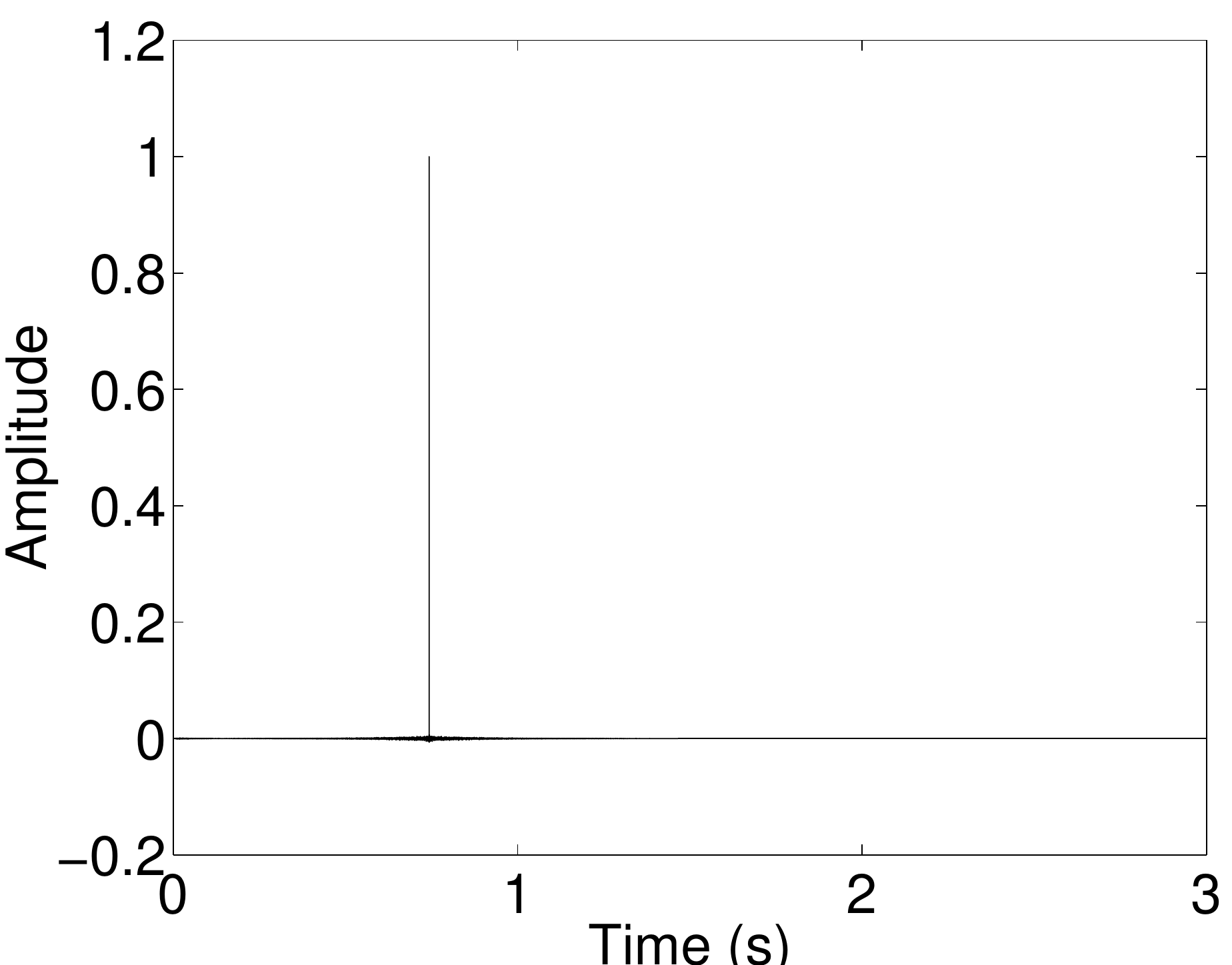}}
\caption{ Inversion process.
  (a), simulated impulse response $\hat{g}(t)$;
  (b), noncausal inverse filter $h(t)$ using regularized inversion;
  (c), shifted and truncated inverse filter $\hat{h}(t)$; and
  (d), dereverberated output $\hat{x}(t)$ ($x(t)=\delta(t)$.}
\label{fig:proc}
\end{figure}

\section{Methods}
\label{sec:methods}

This section describes the design and validation of an open-loop
room response inversion algorithm.  Section~\ref{sec:methods:msrmnt}
describes methods for acquiring validation data (measured room responses)
using impulse-like excitation signals.  Sections~\ref{sec:methods:sim} and \ref{sec:methods:inv} describe methods for simulating
and inverting the room response.

\subsection{Room response measurement}
\label{sec:methods:msrmnt}

Because of the deficiencies described in Section~\ref{sec:back:msrmnt},
few papers in recent decades describe impulse
response measurement techniques using impulse-like excitation signals.
For the application considered in this paper, impulse-like signals have
important advantages. Measurement techniques were therefore developed
to minimize their disadvantages.

\subsubsection{Motivation for the use of starter pistol as an impulse}

Using a starter pistol for room response measurement has three advantages
over non-impulse signals using loudspeakers.
First, measured response need not be deconvolved into an impulse response
because it is already qualitatively similar to the room impulse response.
Second, the SNR is very high because the starter pistol
exceeds 140 dB SPL at 2 m~\cite{MilitaryResearch}.
For a typical background noise level of 50 dB SPL, the SNR 
is 90 dB.
In comparison, the MLS method with order 16 and no repetition has only 60.5 dB
SNR after parameter optimization and compensation for nonlinearities~\cite{Stan2002}.
Therefore, inadequate signal energy is not an issue for a starter pistol.
Third, a starter pistol blast approximates a point source more closely than any
other excitation method considered.
This is good for comparing the measured impulse response
with the simulation from the image source method because the latter assumes
a point source.

Figure~\ref{fig:pist-amb} compares the frequency response of a starter pistol,
estimated as the average of ten pistol shots in an anechoic chamber,
to the frequency response of the ambient noise in a room response
test chamber (a 2 m plywood cube). The ambient noise is 70 dB SPL with
linear weighting, as measured with a Type 2260 B\&K Modular Precision Sound Analyzer.

ISO 3382 specifies a peak SPL at least 45 dB above
the background noise in the frequency range of interest~\cite{ISO-3382}.
Even for a noisy 70 dB SPL environment,
the SNR of a starter pistol shot exceeds 45 dB for the frequency range
280 Hz to 11 kHz, and 30 dB for 110 Hz to 20.5 kHz (Fig.~\ref{fig:pist-amb}).

According to the excitation signal requirements described in Sec.
\ref{sec:back:msrmnt}, the starter pistol still lacks repeatability and
omnidirectionality.  To use it for room response measurement, experimental
methods must be developed to control these two deficiencies.

\subsubsection{Transfer function measurement methods}

Our experiment measures the room impulse response of a 2 m plywood-walled cube.
The cube contains only a microphone and starter pistol;
all other measuring equipment is located outside to avoid any disturbance caused by
obstacles inside the cube. The starter pistol is mounted on the end of a sturdy pipe
and triggered from outside the cube by pulling a cable.

There are two different starter pistol and microphone positions, resulting in a
$2\times2$ matrix transfer function. The exact dimensions of the plywood cube are
$(L_x,\ L_y,\ L_z)$ = (1.84~m, 1.79~m, 1.83~m); table~\ref{tab:positions}
lists the positions of the starter pistol and microphone. Note that the center
of the cube is (0~m, 0~m, 0~m).

Since waveforms of individual pistol shots are not identical (Fig.~\ref{fig:pist-shots}),
we average multiple measurements at the same location.
This repetition has two benefits.
First, we can assume that the averaged impulse response is due to the averaged excitation.
This reduces measurement irregularity, improving repeatability.
Second, SNR improves because the background noise
can be assumed to be independent of room impulse response.

Like any excitation signal, a starter pistol blast is directional.
This variation of signal with respect to angle we label Gun-Related
Transfer Fuction (GRTF).
Figure~\ref{fig:grtf} shows the first 1.5 ms of each GRTF.

\begin{figure}
\centerline{\includegraphics[width=5in]{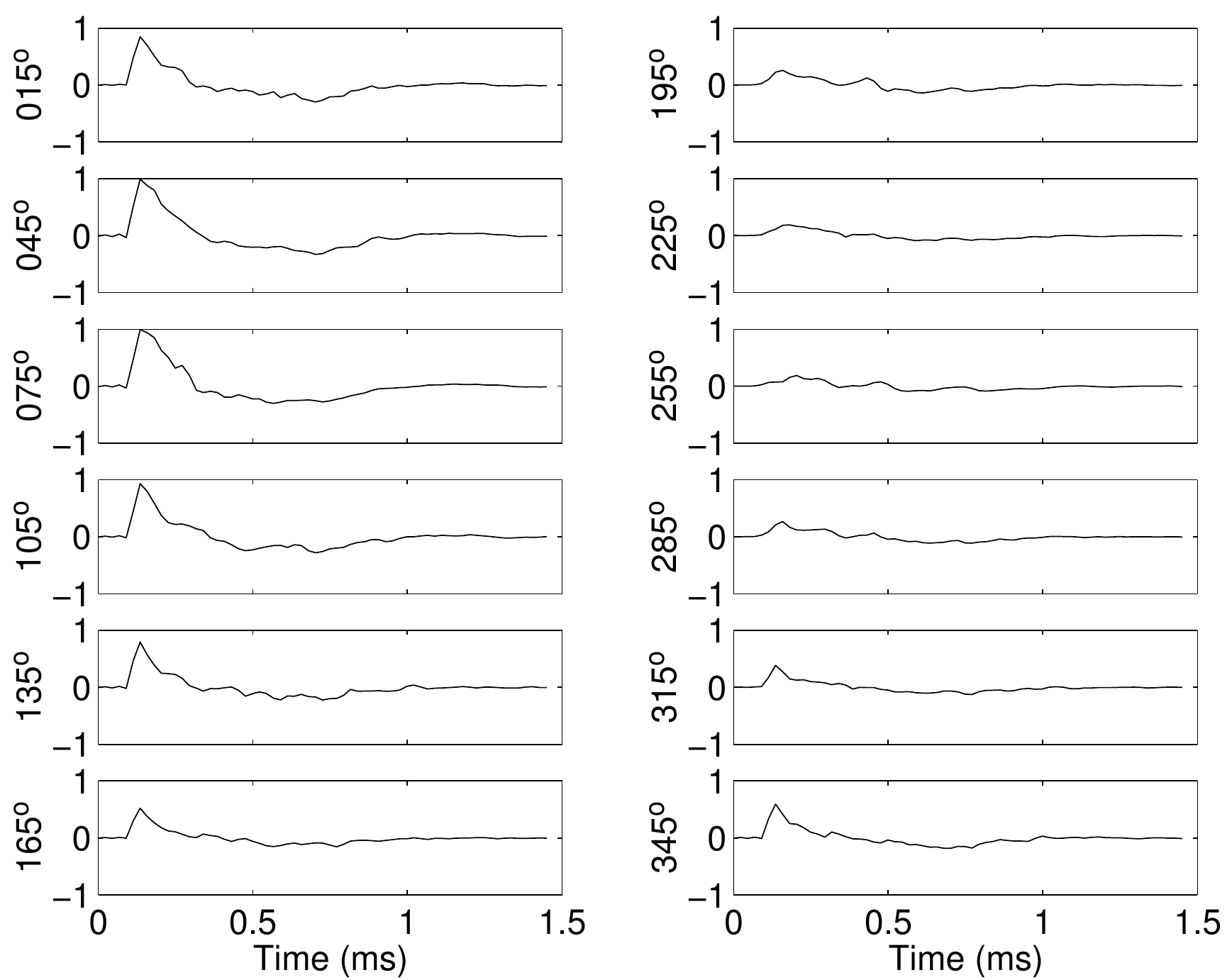}}
\caption{ Gun-Related Transfer Functions (GRTFs).}
\label{fig:grtf}
\end{figure}

Figure~\ref{fig:d-pattern} shows the directional pattern computed from the energy
at each angle.

\begin{figure}
\centerline{\includegraphics[width=4in]{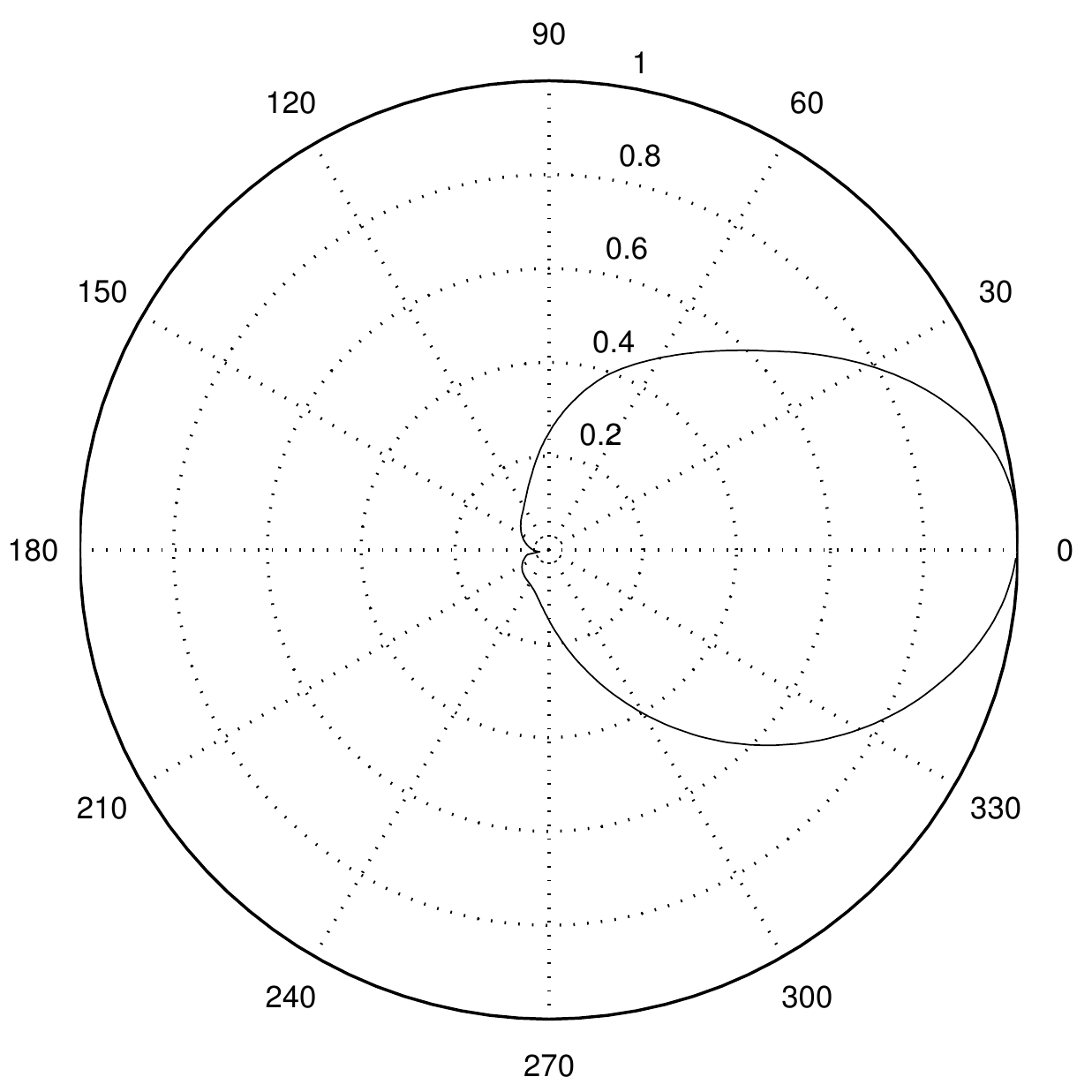}}
\caption{ Directional pattern of the starter pistol blast in normalized linear scale.}
\label{fig:d-pattern}
\end{figure}

To measure the response of the room to an omnidirectional source,
the position of the starter pistol is fixed and the barrel is rotated
to positions $30^\circ$ apart, where $0^\circ$ is directly toward
the microphone, averaging five impulse responses at each angle.
The experiment uses 12 rotation angles, so
a total of 60 shots determine the room impulse response from one point
to another point.

\subsection{Room response simulation}
\label{sec:methods:sim}

It is impractical to directly measure all the room impulse responses
from every loudspeaker position to every mesh point:  only eight loudspeakers
and a $10\times10\times10$ mesh demand 8000 experiments.  Instead,
inverse transfer functions are derived from
approximate room responses simulated with the image source method.
This then demands verification of the simulation, using error
metrics to compare corresponding pairs of measured and simulated responses.

Parameters of the image source simulation include the room
dimensions, the position of the sound source and receivers, the
speed of sound (dependent on temperature and relative
humidity\cite{Cramer1993}), and the reflection coefficient $r$ of the
wall material (Eq.~(\ref{eq:ism})).  Although $r$ varies with frequency,
modeling the frequency dependence greatly increases the computational
cost of the simulation.  All simulations reported in this paper
therefore assume a frequency-independent $r$, related to the
average Sabine absorptivity $\bar{a}$:~\cite{Allen1979}
\begin{equation*}
\bar{a}=1-r^2
\end{equation*}
The value of $\bar{a}$ is optimized experimentally.

The simulation uses a sampling frequency of 44.1 kHz. The
length of the simulated impulse response is 65536 samples, about 1.5
seconds.  The speed of sound is taken to be 346.58 m/s, based on
Cramer's equation evaluated at the temperature and relative humidity
measured in the plywood cube ($24.4^\circ$C$, 37.5\%$)~\cite{Cramer1993}.

Room response simulations were evaluated using
three metrics: local mean-squared error (described here), global and local
dereverberation ratios (described in the next section), and remainder
reverberation time (described in the next section). 
Mean-squared error measures time-domain similarity, i.e., how
alike the amplitude-versus-time graphs of the two responses look.
For an $M$ sample interval starting at the $k^{th}$ sample,
this error is expressed as
\begin{equation}
E_{ms}[k] = \frac{1}{M} \sum_{n=k}^{k+M-1}
            \left(\frac{\hat{g}[n]}{\hat{g}_{rms}}-\frac{g[n]}{g_{rms}}\right)^2
\label{eq:mse}
\end{equation}
where $\hat{g}[n]$ and $g[n]$ are the simulation and measurement of the room
impulse responses, and $\hat{g}_{rms}$ and $g_{rms}$ are their RMS values
in the interval $[k, k+M-1]$.

For an actual room response $G_{orig}(z)$ and an 
excitation signal spectrum $S(z)$, the measured room response is
$G(z)=S(z)G_{orig}(z)$.  When the excitation signal is a starter pistol,
$S(z)$ may be measured by recording the pistol impulse response $s(t)$
in an anechoic chamber; when the excitation signal is pseudo-noise or a
sine sweep, $S(z)$ must be computed by multiplying the theoretical
pseudo-noise spectrum with the loudspeaker frequency response.
Pseudo-noise room response measurement techniques may then compare
$G(z)S(z)^{-1}$, the source-corrected room response, with $\hat{G}(z)$,
the simulated room response.  When $S(z)$ is the spectrum of a starter
pistol, however, $S(z)^{-1}$ has undesirable properties
(it is high-pass, noncausal, and nearly singular), so
the ``source-corrected'' room response $g(t)\ast s(t)^{-1}$ is difficult
to evaluate visually.  Conversely, because $s(t)$ is impulse-like,
visual comparison of the measured response $g(t)$ with $\hat{g}(t)\ast
s(t)$ (the measured excitation filtered by simulated response) is
natural and meaningful.

\begin{figure}
\centerline{\includegraphics[width=3.5in]{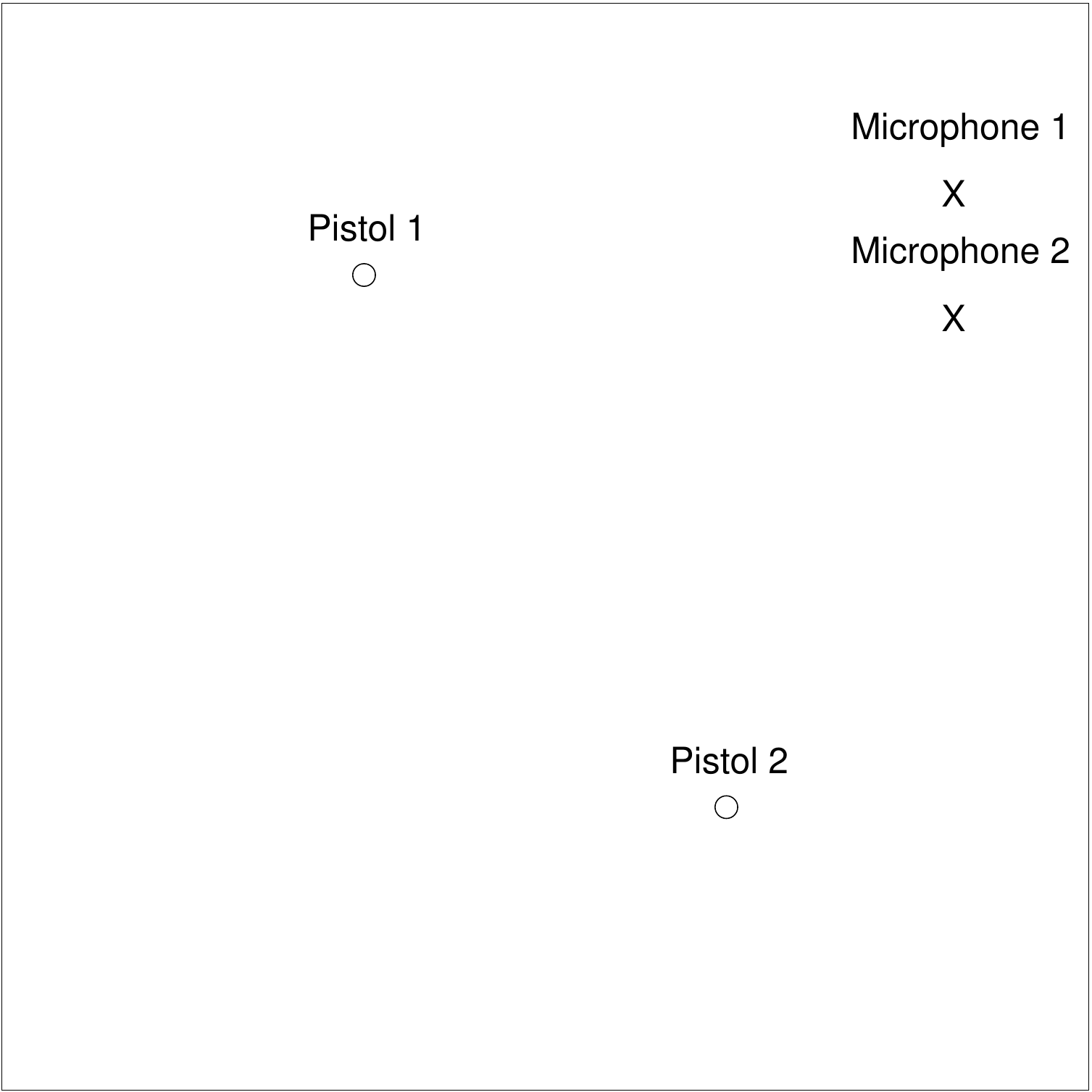}}
\caption{ Experiment setup.}
\label{fig:setup}
\end{figure}

Figure~\ref{fig:first20ms} compares the first 20 ms of two impulse responses
from starter pistol position 2 to microphone position 1 (see Fig.
\ref{fig:setup} and table~\ref{tab:positions}).
The upper plot shows $g(t)$, the average of
the 60 measured impulse responses. The lower plot is
$\hat{g}(t) \ast s(t)$, where $\hat{g}(t)$ is computed using the
image source method. The very close match between these two impulse
responses validates both the image source method and the angle-averaged
pistol measurement.

\begin{figure}
\centerline{\includegraphics[width=5in]{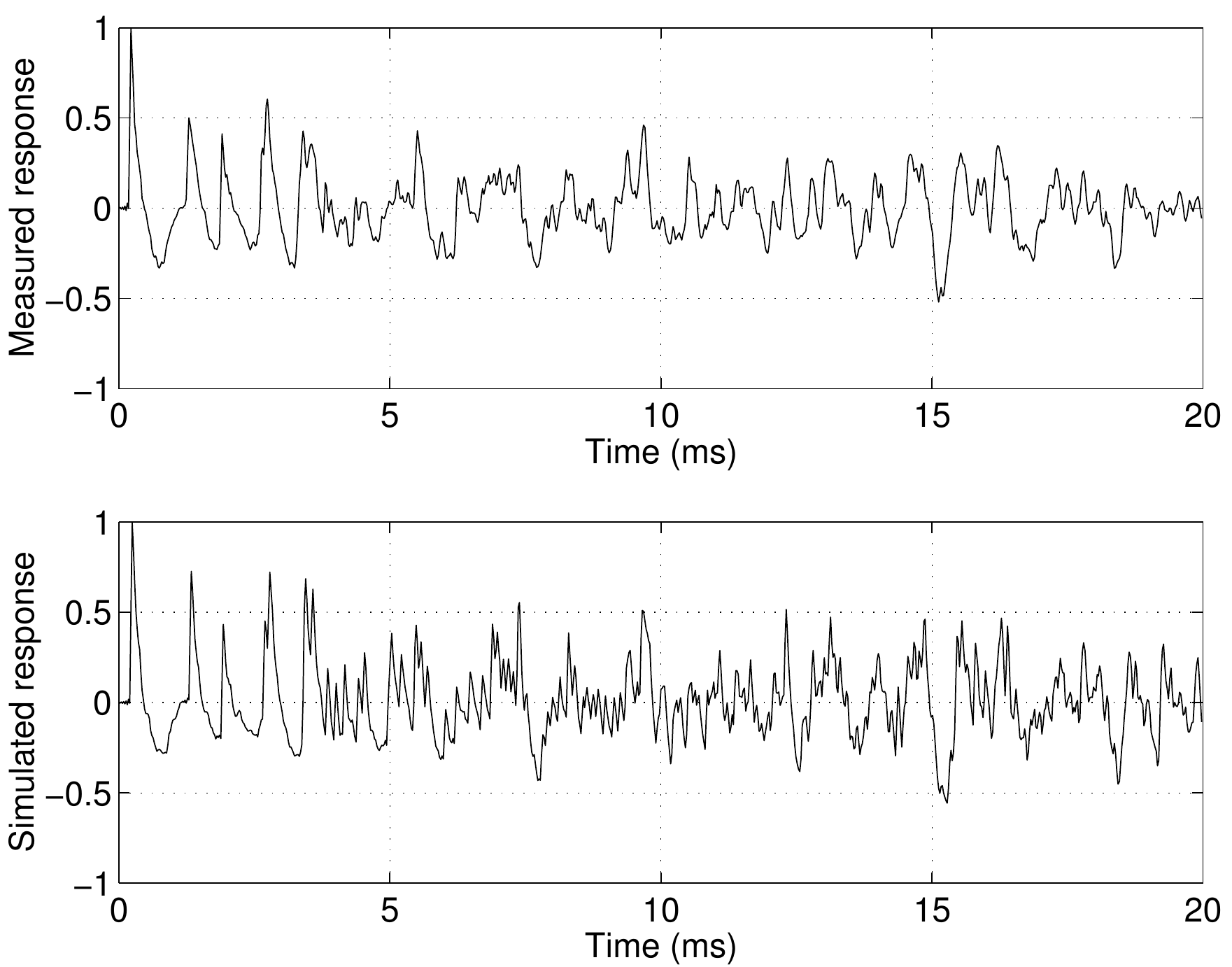}}
\caption{ First 20 ms of measured and simulated impulse responses.}
\label{fig:first20ms}
\end{figure}

The simulation preserves the peak locations closely 
even after 100 ms (Fig.~\ref{fig:after100ms}), but the visual similarity
of the signals is not as great as during the first 20 ms
(Fig.~\ref{fig:first20ms}).

\begin{figure}
\centerline{\includegraphics[width=5in]{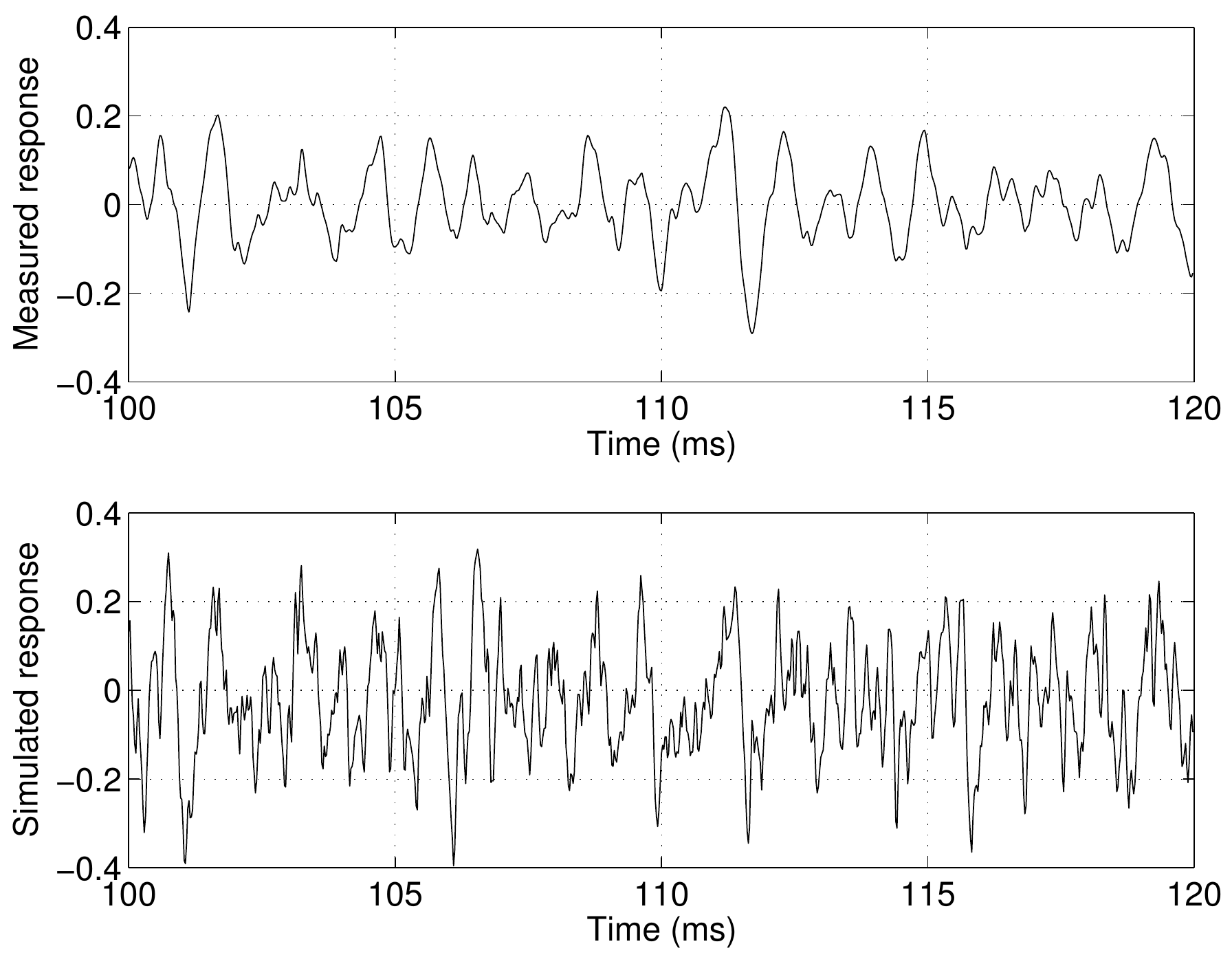}}
\caption{ Measured and simulated impulse responses after 100 ms.}
\label{fig:after100ms}
\end{figure}

The increasing dissimilarity between $g(t)$ and $\hat{g}(t)\ast s(t)$ as
$t$ increases is quantified by a gradual increase in the local
mean-squared error of the simulation, computed using Eq.~(\ref{eq:mse})
with intervals of $20$ and $100$ ms (Fig.~\ref{fig:mse}). 
This time-dependent dissimilarity may be explained by considering the
accumulated effect on $g(t)$ of frequency-dependent wall reflections
and air propagation filtering.  $\hat{g}(t)$ is computed using the
time-domain image source method, which does not model frequency-dependent
wall reflections and air propagation.

\begin{figure}
\centerline{\includegraphics[width=5.1in]{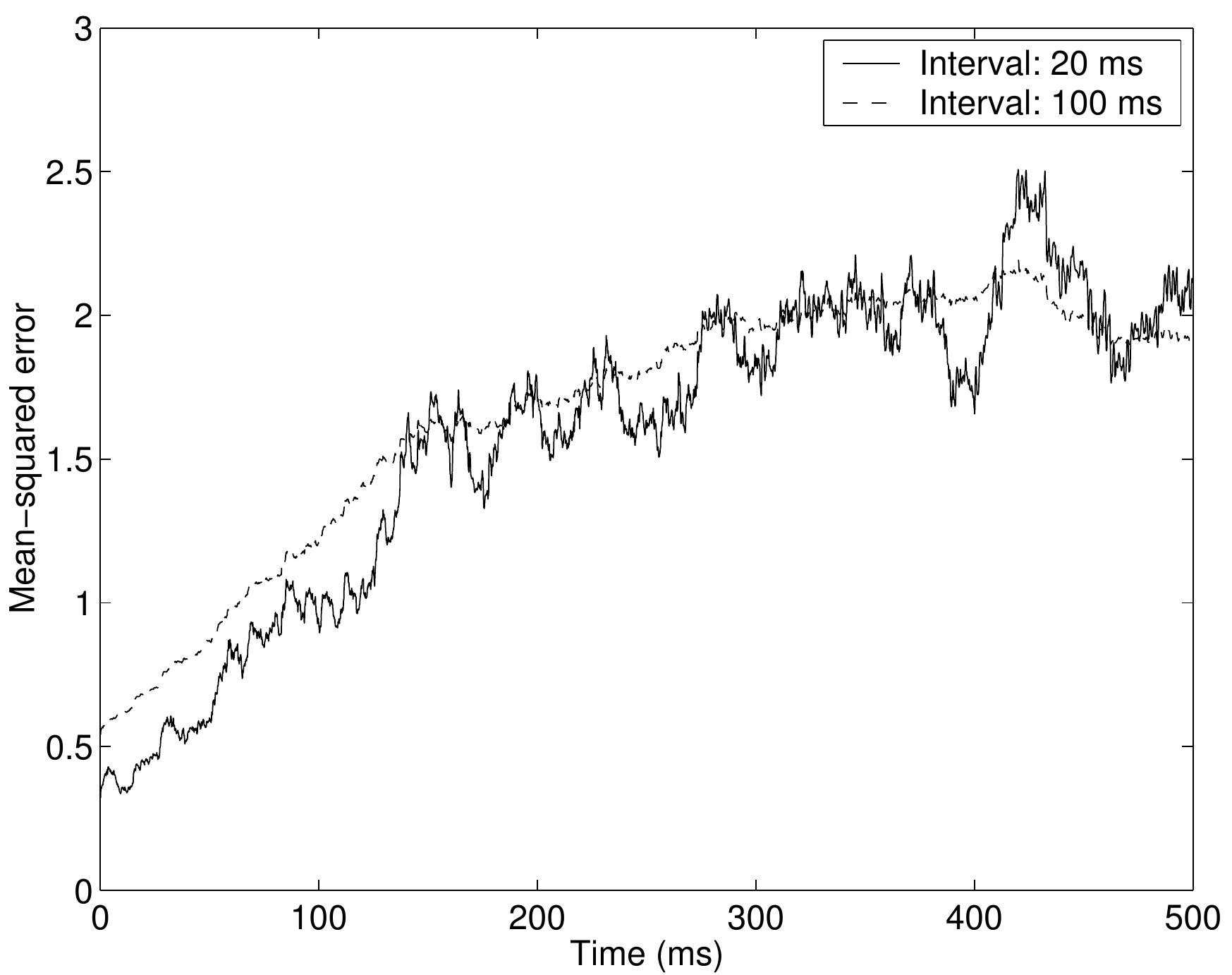}}
\caption{ Mean-squared error between simulation and measurement.}
\label{fig:mse}
\end{figure}

Figure~\ref{fig:ply_attn} shows the filtering effect of wall reflections
on the spectrum of a single acoustic ray. After only one reflection,
frequencies below 10 Hz are attenuated 20 dB relative to frequencies
above 100~Hz~\cite{Kinsler2000}. After ten reflections (70 to 100 ms),
frequencies below 100 Hz are effectively zeroed.
Figure~\ref{fig:air_attn} shows the filtering effect of propagation
through air at $20^\circ$ C and 30\% relative humidity~\cite{CRC_Handbook}.
After 34.3 m (100 ms), spectral components at 10 kHz are attenuated
about 8 dB.
The attenuation due to wall reflections and air propagation is enormous
even after 100 ms.

\begin{figure}
\centerline{\includegraphics[width=5.1in]{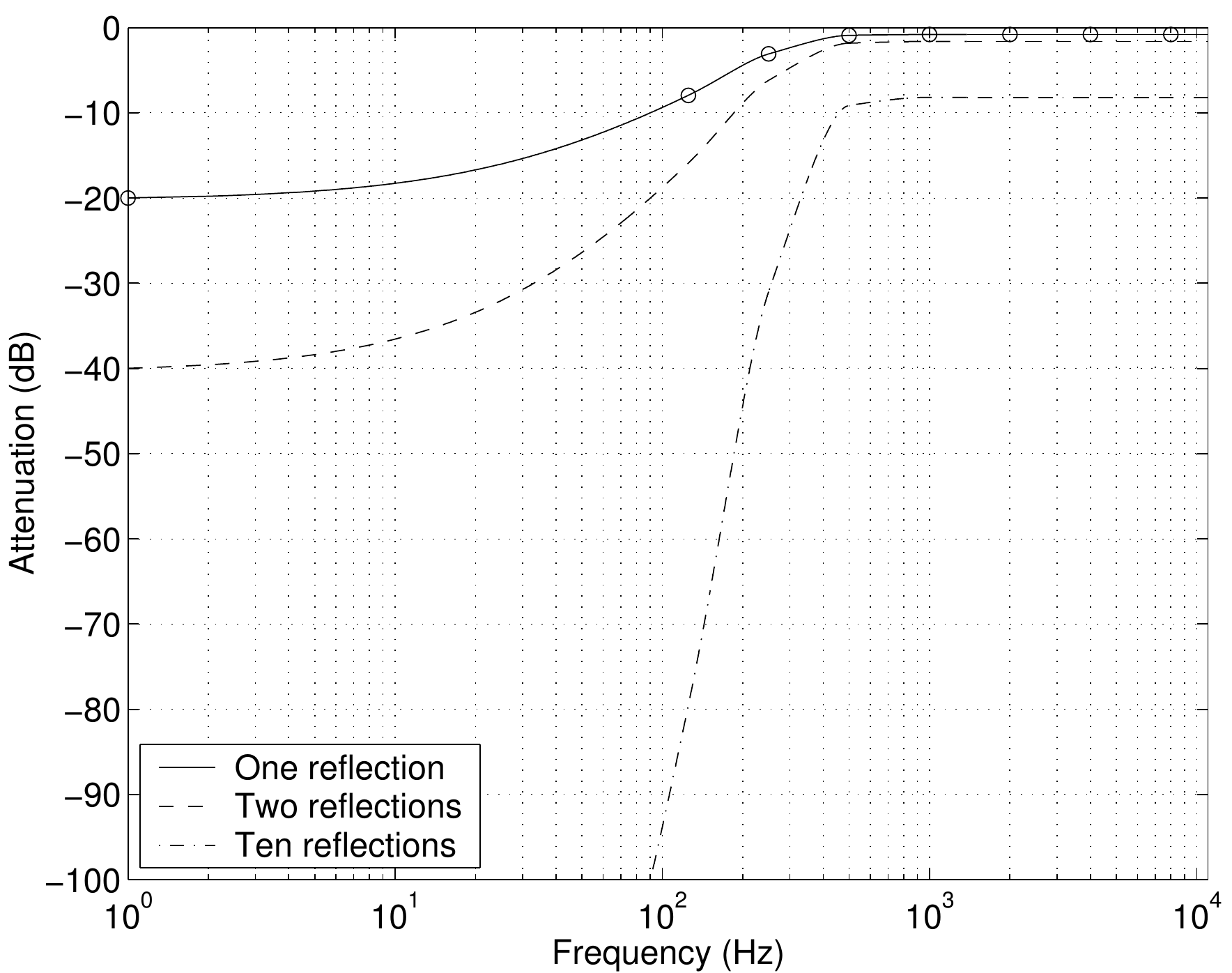}}
\caption{ Frequency response after one, two, and ten wall reflections.  (Circled data points are adapted from \cite{Kinsler2000}.)}
\label{fig:ply_attn}
\end{figure}

\begin{figure}
\centerline{\includegraphics[width=4.8in]{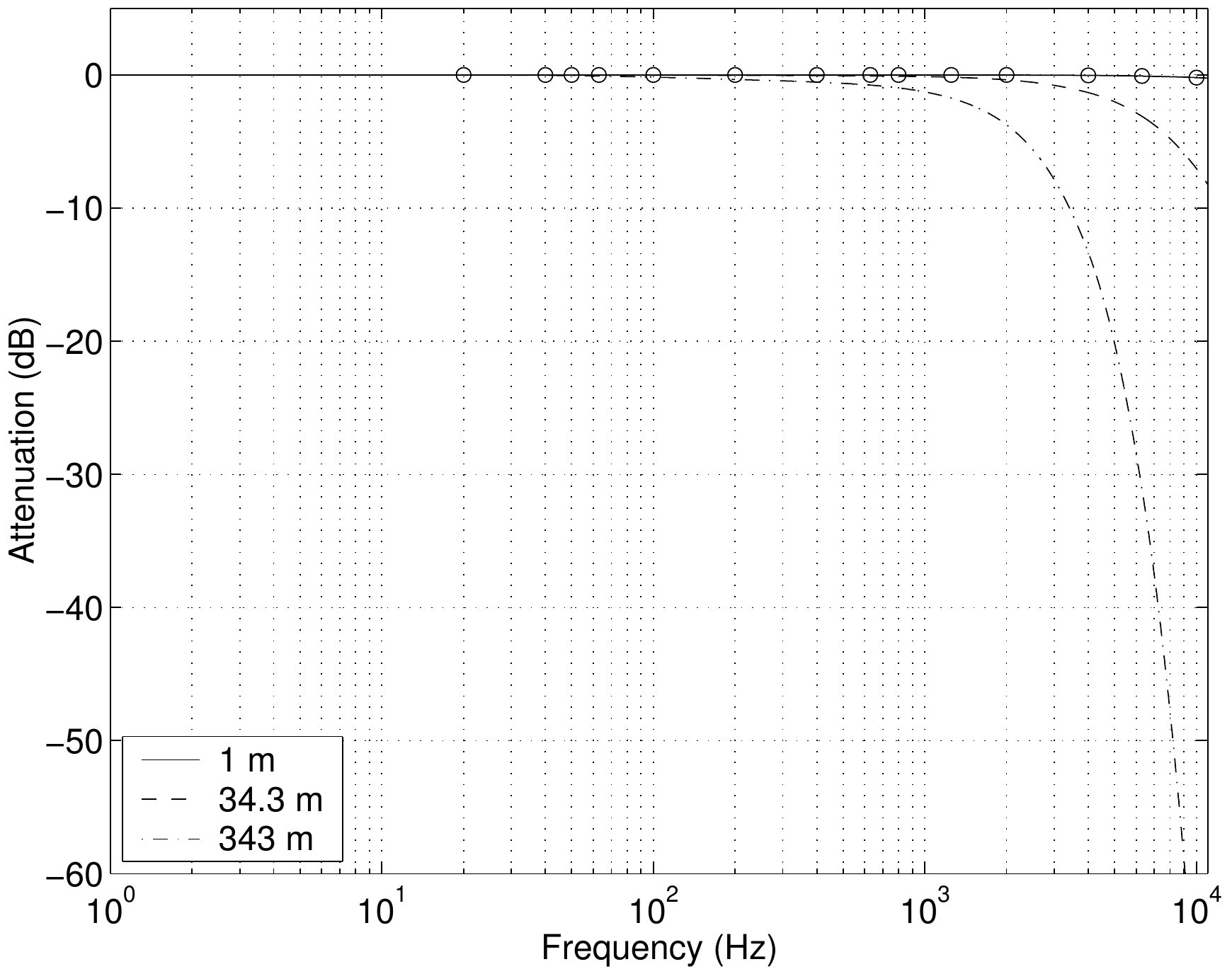}}
\caption{ Frequency response of propagation through air at $20 \deg$ C, 30\% relative humidity after 1 m, 34.3 m, and 343 m.  (Circled data points are adapted from \cite{CRC_Handbook}.)}
\label{fig:air_attn}
\end{figure}

\subsection{Room response inversion}
\label{sec:methods:inv}

Room responses simulated using the image source method are next inverted
using the method of regularized inversion with modeling
delay~\cite{Kirkeby1998}.  Experiments indicate that effective
inversion requires $\beta > 0$, but that the exact value of $\beta$ in
the range $10^{-4}\le\beta\le 1$ has little effect on inversion
performance.
The value of the modeling delay $D$ is more important. Inversion results
improve almost monotonically as $D$ increases, suggesting that $D$
be the longest modeling delay imperceptible to users.
Results in this paper use $\beta=10^{-2}$ and $D=500$ ms. 
Finally, Eq.~(\ref{eq:reginv}) may be used for either scalar inversion
(one speaker, one control point) or matrix room
response inversion ($L$ speakers, $M$ control points).  This paper
reports results of both scalar and matrix inversion experiments, where
matrix inversion is performed with $L=2$, $M=2$ using the geometry shown
in Fig.~\ref{fig:setup} and table~\ref{tab:positions}.

\subsubsection{Evaluation of room impulse response inversion}

Room response inversion can eliminate the perceptual ``signature''
of a room by attenuating early echoes; it can also reduce long-term
reverberant energy.

The early echoes should be well suppressed because they characterize
the perceived geometry of the room. The later portion of the room response
is related more to wall material and room size than to specific room
geometry.

Assuming that the desired signal $x(t)$ at a certain
location is an impulse, the output $\hat{x}(t)$ at that location 
needs to be as close to an impulse as possible: it should
contain as little energy as possible at time $t \neq 0$. 
The output is expressed as $\hat{X} = G\hat{H}X$ where $G$ is a measured
impulse response and $\hat{H}$ is the approximate inverse filter created
from the simulation $\hat{G}$ using Eqs.~(\ref{eq:moddelay}),
(\ref{eq:trunc}), and (\ref{eq:reginv}).
The time-domain expression of the output is
\begin{equation*}
\hat{x}(t) = g(t) \ast \hat{h}(t) \ast  x(t)
\end{equation*}
or, for matrix inversion experiments,
\begin{equation*}
\hat{x}_k(t)=\sum_{j=1}^L\sum_{i=1}^Mg_{kj}(t)\ast
             \hat{h}_{ji}(t) \ast x_i(t)
\end{equation*}

To claim that the inverse filter dereverberates the room impulse response,
for an input $x(t)=\delta(t)$, the filtered output
$\hat{x}(t)=\hat{h}(t) \ast g(t)$ should be similar to a delayed impulse,
$\hat{x}(t) \approx \delta(t-D)$.
The success of dereverberation may be measured by computing the residual
energy in the signal $\hat{x}(t)$ at times $|t-D| > T_{min}$, for some small
value of $T_{min}$. The residual energy in $\hat{x}(t)$ is computed as
\begin{equation*}
E_{resid}(\infty) = \int_{T_{min}<|t-D|}\hat{x}^2(t)dt
\end{equation*}
``Early echoes'' may be defined to be causal or noncausal echoes within a time
window $|t-D|<T$. The residual energy within $T$ seconds is computed as
\begin{equation*}
E_{resid}(T) = \int_{T_{min}<|t-D|<T}\hat{x}^2(t)dt
\end{equation*}
The efficacy of the dereverberation is described by the
``dereverberation ratio'' (DR): the ratio of the original room response energy
$\int_{T_{min}}^{T} g^2(t)$ to the residual energy, thus,
\begin{equation*}
DR(\infty)= 10\log_{10}\frac{\int_{T_{min}}^{\infty} g^2(t) dt}
                            {\int_{T_{min}<|t-D|}\hat{x}^2(t)dt}
\end{equation*}
\begin{equation*}
DR(T) = 10\log_{10}\frac{\int_{T_{min}}^{T} g^2(t) dt}
                        {\int_{T_{min}<|t-D|<T}\hat{x}^2(t)dt}
\end{equation*}
For the output $\hat{x}(t)$ to have less energy than the
measured impulse response $g(t)$, both decibel ratios should be
positive. They can therefore be used to evaluate
and optimize the simulation and inversion of the room impulse responses.

Simulation and inversion can also be evaluated using the remainder
reverberation time $T_L$, defined implicitly as
\begin{equation*}
L = 10\log_{10}\frac{\int_{0}^\infty g^2(t)dt}
                    {\int_{T_L}^{\infty}\hat{x}^2(t)dt}
\end{equation*}

The remainder reverberation times $T_{10}$, $T_{20}$, and $T_{60}$
of both measured and dereverberated outputs will be compared.
The reference for the remainder reverberation time is the integrated
energy of the measured room impulse response $g(t)$.

\subsubsection{Optimization of the window for for impulse response inversion}

An inverse filter created using a complete 1.5 second simulation
of the room response fails:  the energy of the
dereverberated output exceeds the energy of the measured impulse
response.

The mean-squared error with respect to time ($E_{ms}$) indicates
that the accuracy of the
image source simulation decreases with time (Fig.~\ref{fig:mse}).
This suggests that the dereverberation ratio may improve by applying
a tapering window, such as an exponential with time constant $\tau$: 
\begin{equation*}
\tilde{g}(t) = e^{-t/\tau}\hat{g}(t)
\end{equation*}

Figure~\ref{fig:DR_tau} shows the dereverberation ratios $DR(\infty)$
and $DR(100$ ms) for both scalar and matrix inversion, using
$\tilde{g}(t)$ instead of $\hat{g}(t)$ in order to create the inverse filter,
with values of $\tau$ = 0.01, 0.02, 0.04, 0.08, 0.16, 0.32, and 0.64 s.

\begin{figure}
\centerline{\includegraphics[width=4.5in]{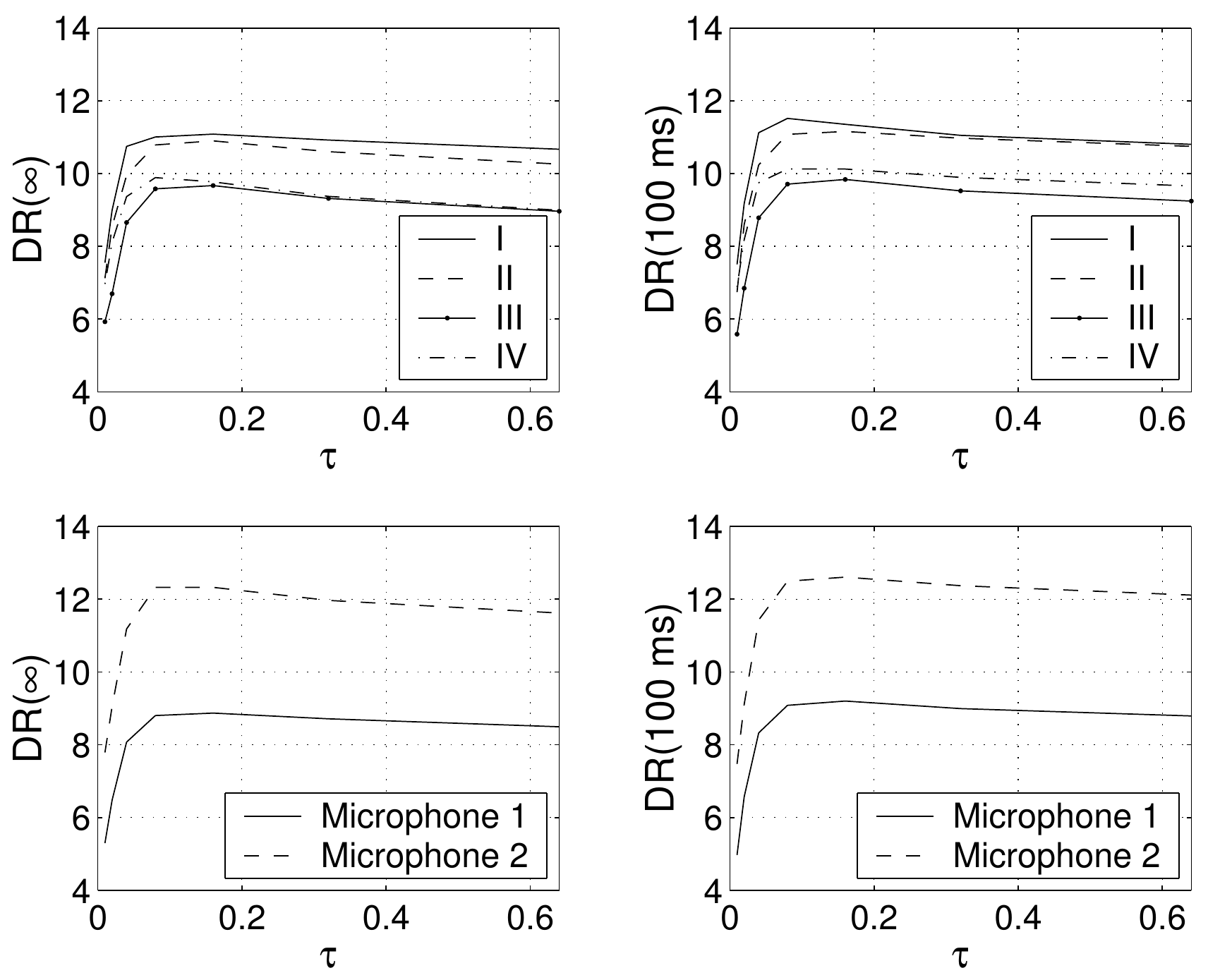}}
\caption{ Dereverberation ratios with respect to the time constant.  Upper graphs, scalar inversion; lower graphs, matrix inversion.}
\label{fig:DR_tau}
\end{figure}

According to the dereverberation ratios depicted above, 
$\tau=0.06$ s is close to optimal.
\section{Results}
\label{sec:results}

Two measures are used to discuss the inversion results.
First, the total and the early dereverberation ratios
$DR(\infty)$ and $DR(T)$ are compared.
Second, the remainder reverberation times $T_{10}$,
$T_{20}$, and $T_{60}$ of measured and dereverberated
responses are compared.

Results of scalar inversion for four different impulse responses and of
matrix inversion of a $2 \times 2$ matrix transfer function are presented
here.  The four different room response geometries used for
both scalar and matrix inversion may be numbered as follows, with
reference to Fig.~\ref{fig:setup} and table~\ref{tab:positions}: 
Impulse response number I, pistol 1 to microphone 1; 
II, pistol 1 to microphone 2; III, pistol 2 to microphone 1;
and IV, pistol 2 to microphone 2. For $2 \times 2$ matrix inversion,
room responses at microphone location 1 and 2 are called microphone 1 and
microphone 2 respectively.

\begin{table}
\center
\caption{\label{tab:positions}Positions (in meters) of pistol and microphone in plywood cube.}
\begin{tabular}{cccc}
\hlinewd{2pt}
           & Position 1                & Position 2                \\\hline
Pistol     & $( 0.26,\  0.30,\ -0.15)$ & $(-0.26,\ -0.30,\ -0.15)$ \\
Microphone & $(-0.57,\  0.58,\  0.31)$ & $(-0.39,\  0.58,\  0.31)$ \\
\hlinewd{2pt}
\end{tabular}
\end{table}

Section~\ref{sec:results:absorp} describes
experiments designed to optimize and validate the image source method. 
Section~\ref{sec:results:ind} describes scalar inversion results with and
without windowing the simulated room response. Section~\ref{sec:results:mat}
describes $2 \times 2$ matrix inversion results with windowing.

\subsection{Optimization of absorption coefficient}
\label{sec:results:absorp}

When all interior surfaces of the room are covered with the same material
and the reverberation time is known, average Sabine absorptivity $\bar{a}$
is given directly by Sabine's formula
\begin{equation*}
\bar{a} = \frac{0.161V}{ST_{60}}
\end{equation*}
where $V$ and $S$ are the volume and the surface area of the room
respectively~\cite{Kinsler2000}.

The measured $T_{60}$ of the 2 m plywood cube using Schroeder's
integration formula \cite{Schroeder1965} is 1.32 s, yielding
$\bar{a}= 0.0407$.
Since this is merely an estimation from
Sabine's formula, it was bracketed with values 
0.01, 0.02, 0.04, 0.08, and 0.16.

Figure~\ref{fig:DR-absorp} shows that dereverberation ratios 
are not affected by the modeled absorption coefficient $\bar{a}$.
This indicates that the phase information of the room impulse response
is more important than the magnitude information, i.e., for inversion
the exact timing of the reflections is more important than their magnitudes.

\begin{figure}
\centerline{\includegraphics[width=4.5in]{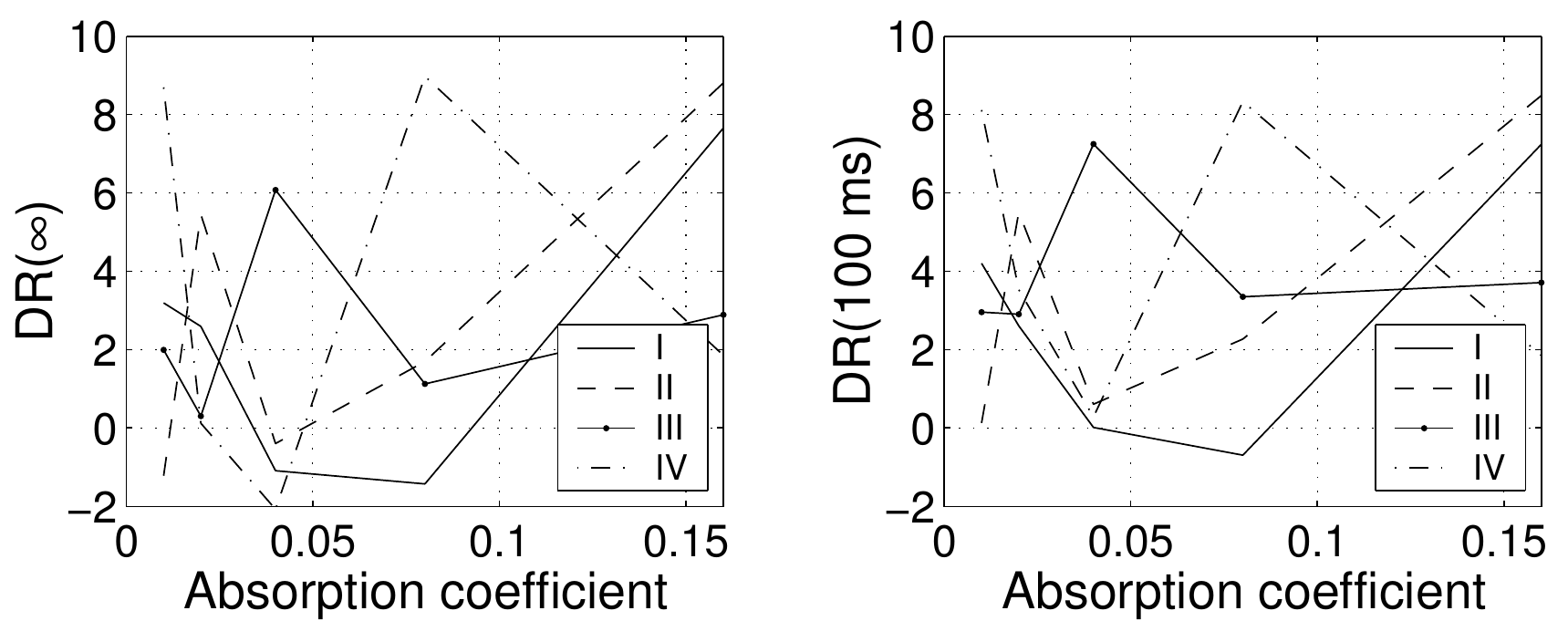}}
\caption{ Dereverberation ratios with respect to absorption coefficient $\bar{a}$.}
\label{fig:DR-absorp}
\end{figure}

\subsection{Scalar inversion with and without windowing}
\label{sec:results:ind}

Scalar inversion was performed both without and with exponential windowing
of the simulated response.
Figure~\ref{fig:out_lin_nowin} depicts the measured impulse response and
the response ``dereverberated'' without windowing
from starter pistol location 1 to microphone location 2
(impulse response number II).
Table~\ref{tab:DR_ind} shows that the supposedly dereverberated
impulse responses have approximately 10 dB more energy than the measured impulse
responses; so dereverberation fails when no
tapering window is applied to the simulated impulse reponses.

\begin{figure}
\centerline{\includegraphics[width=4.5in]{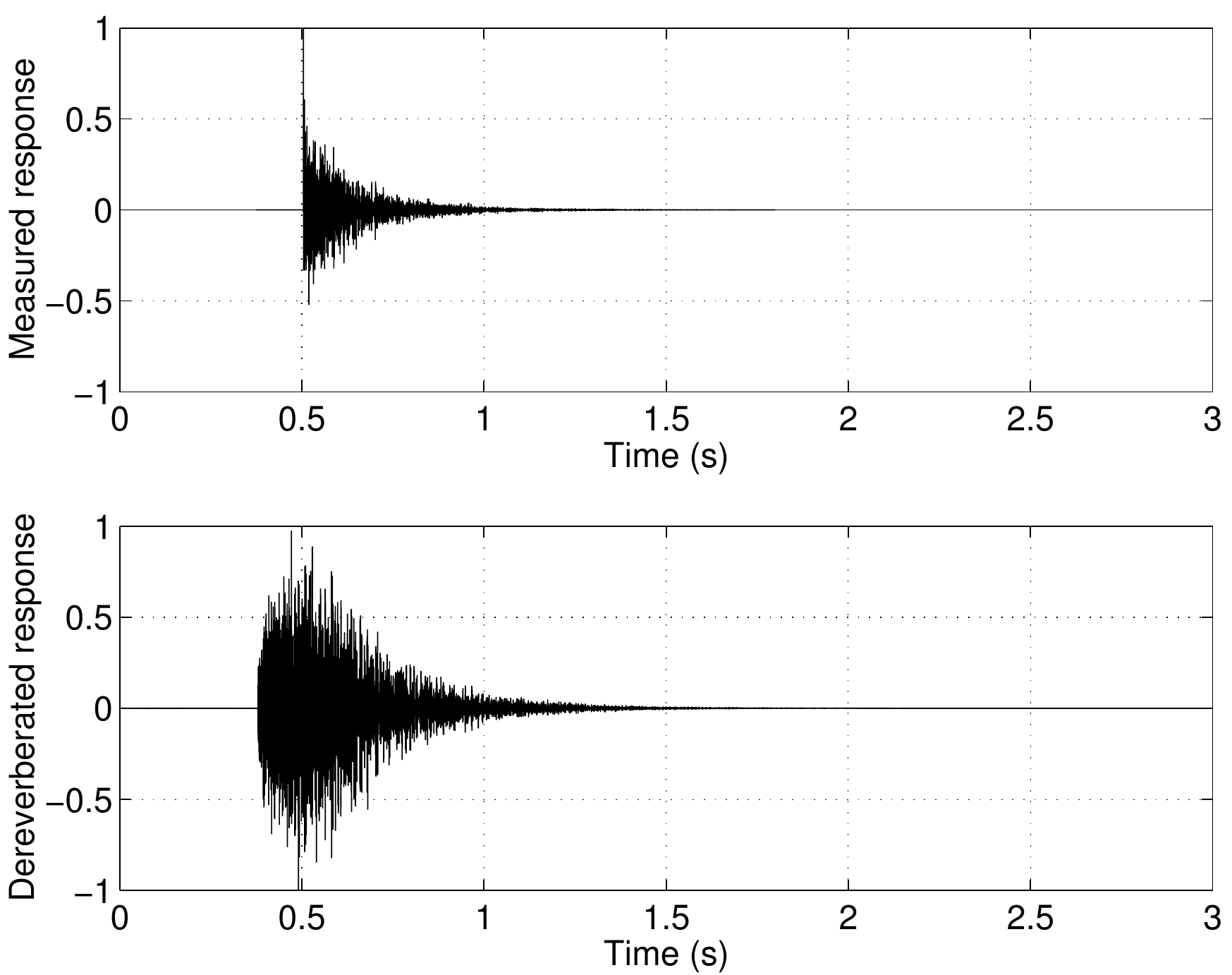}}
\caption{ Linear plots of measured and dereverberated impulse responses (scalar inversion; impulse response number III).  No windowing is applied to the simulated impulse response $\hat{g}(t)$.}
\label{fig:out_lin_nowin}
\end{figure}

Figure~\ref{fig:out_ind} depicts the results of
dereverberation, where simulated impulse responses are windowed by
an exponential window with time constant $\tau=0.06$ s.
Table~\ref{tab:DR_ind} shows that
the dereverberated impulse responses have from $8.11$ to $10.98$ dB
less energy than the measured impulse responses; dereverberation works.

\begin{figure}
\centerline{\includegraphics[width=4.5in]{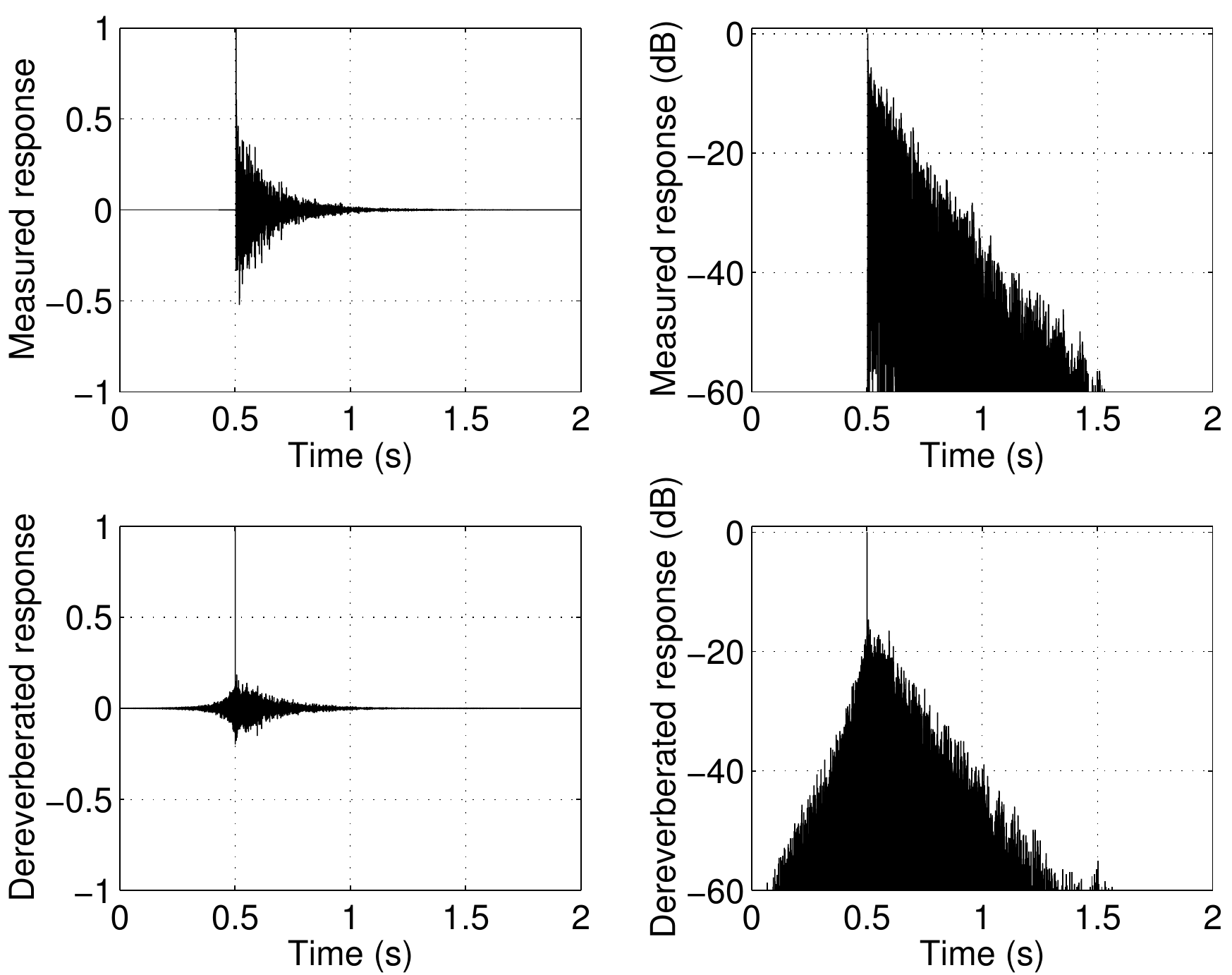}}
\caption{ Measured and dereverberated impulse responses (scalar inversion; impulse response number III).  An exponential window with time constant $\tau=0.06$ s is applied to the simulated impulse response $\hat{g}(t)$.}
\label{fig:out_ind}
\end{figure}

\begin{table}
\center
\caption{\label{tab:DR_ind}Dereverberation ratios in dB of dereverberated impulse responses.}
\begin{tabular}{ccccc}
\hlinewd{2pt}
    &\multicolumn{2}{c}{Non-windowed}&\multicolumn{2}{c}{Windowed} \\
    & $DR(\infty)$ & $DR(100$ ms$)$ & $DR(\infty)$ & $DR(100$ ms$)$ \\\hline 
I   & $-8.95$      & $-9.03$      & $10.98$      & $11.45$         \\
II  & $-10.54$     & $-9.85$      & $10.56$      & $10.84$         \\ 
III & $-10.11$     & $-10.04$     & $8.11$       & $8.30$          \\
IV  & $-8.88$      & $-8.60$      & $10.56$      & $10.84$         \\
\hlinewd{2pt}
\end{tabular}
\end{table}

Figure~\ref{fig:int_dec} shows integrated energy decay curves of four
different dereverberated impulse responses with respect to the measured
impulse resposes; table~\ref{tab:rt_ind} lists the dereverberation
times of scalar inversion according to these decay curves.

\begin{figure}
\centerline{\includegraphics[width=4.5in]{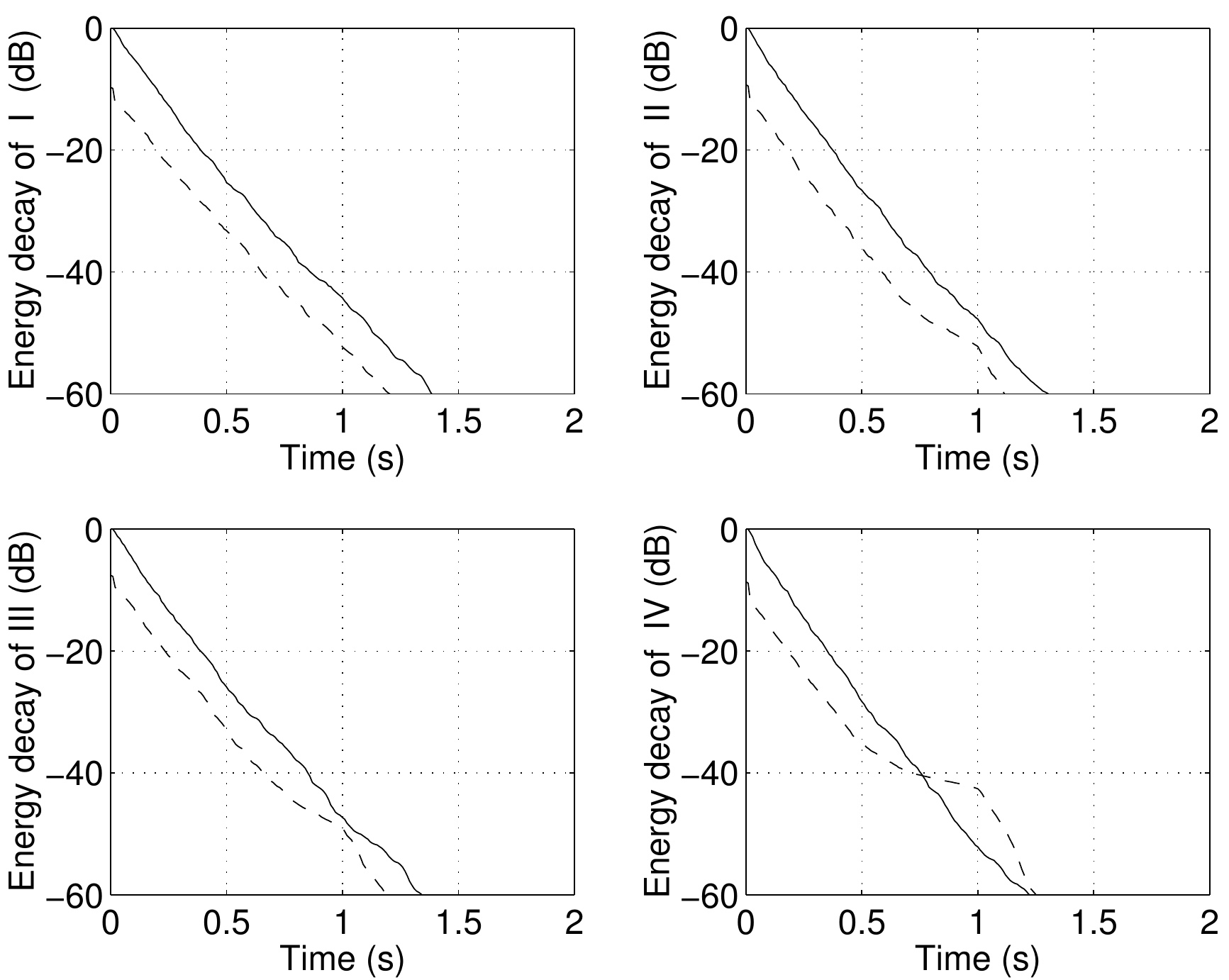}}
\caption{ Integrated energy decay curves for scalar inversion.  Solid line is measured response; dashed line is dereverberated response.}
\label{fig:int_dec}
\end{figure}

\begin{table}
\center
\caption{\label{tab:rt_ind}Remainder reverberation times $T_{10}$, $T_{20}$, and $T_{60}$, in seconds, of measured and dereverberated impulse responses; scalar inversion.}
\begin{tabular}{ccccccc}
\hlinewd{2pt}
    & \multicolumn{3}{c}{Measured responses} & \multicolumn{3}{c}{Dereverberated responses} \\  
    & $T_{10}$ & $T_{20}$ & $T_{60}$ & $T_{10}$ & $T_{20}$ & $T_{60}$ \\\hline 
I   & $0.20$   & $0.39$   & $1.39$   & $0.02$   & $0.20$   & $1.21$   \\
II  & $0.18$   & $0.38$   & $1.31$   & $0.02$   & $0.18$   & $1.12$   \\ 
III & $0.19$   & $0.39$   & $1.35$   & $0.04$   & $0.24$   & $1.19$   \\
IV  & $0.18$   & $0.36$   & $1.23$   & $0.02$   & $0.19$   & $1.26$   \\
\hlinewd{2pt}
\end{tabular}
\end{table}

\subsection{Matrix inversion with windowed simulation}
\label{sec:results:mat}
Figures~\ref{fig:out_mat_lin} and \ref{fig:out_mat_dB} depict $2 \times 2$ matrix
inversion results; dereverberation ratios are in table~\ref{tab:DR_mat}.
Remainder reverberation energy decay curve and corresponding remainder
reverberation times ($T_{10}$, $T_{20}$, and $T_{60}$) are shown
in Fig.~\ref{fig:dec_mat} and table~\ref{tab:rt_mat}.

\begin{figure}
\centerline{\includegraphics[width=4.5in]{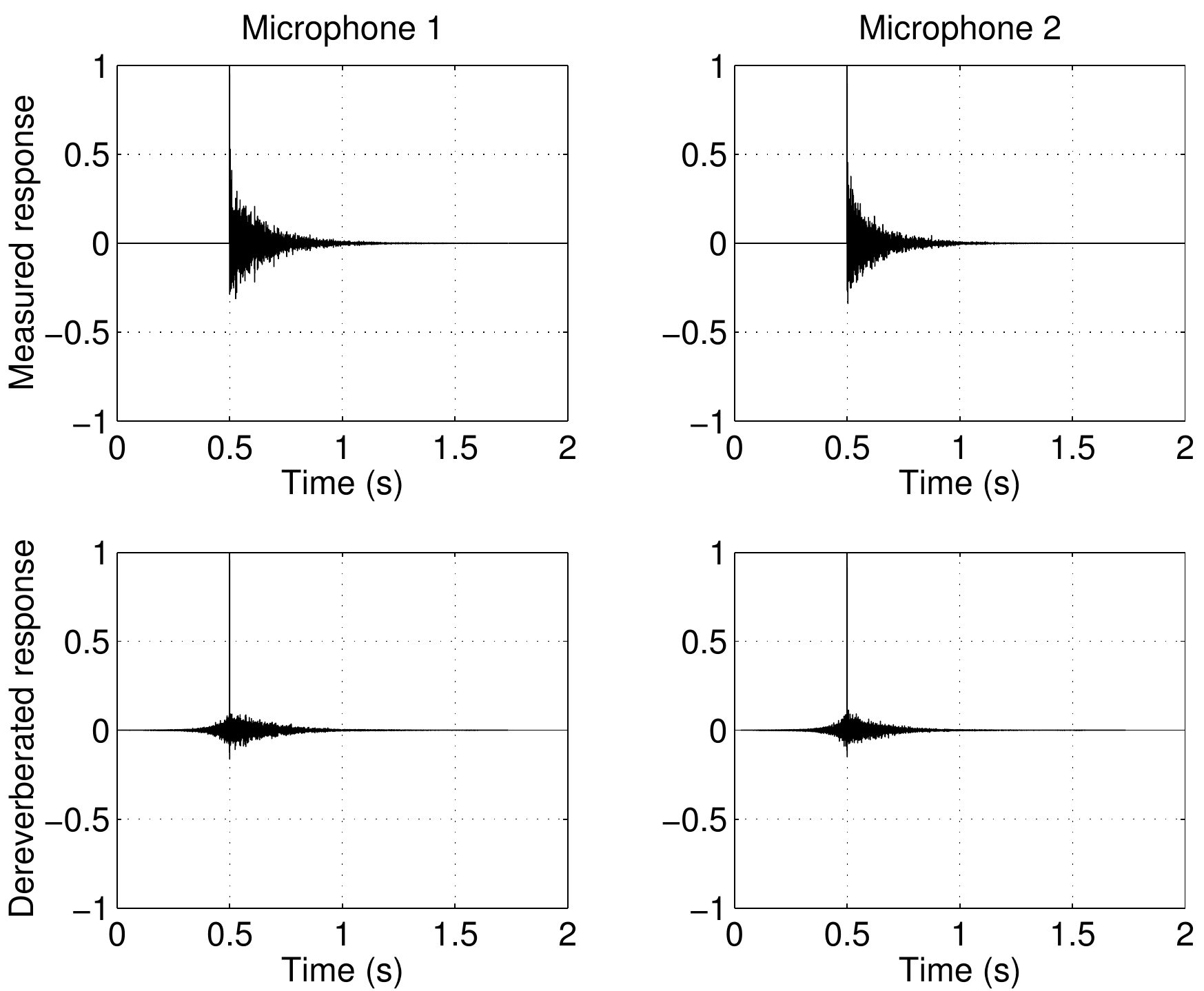}}
\caption{ Linear plots of measured and dereverberated impulse responses ($2 \times 2$ matrix inversion).  An exponential window with time constant $\tau=0.06$ s is applied to the simulated impulse response $\hat{g}(t)$.}
\label{fig:out_mat_lin}
\end{figure}

\begin{figure}
\centerline{\includegraphics[width=4.5in]{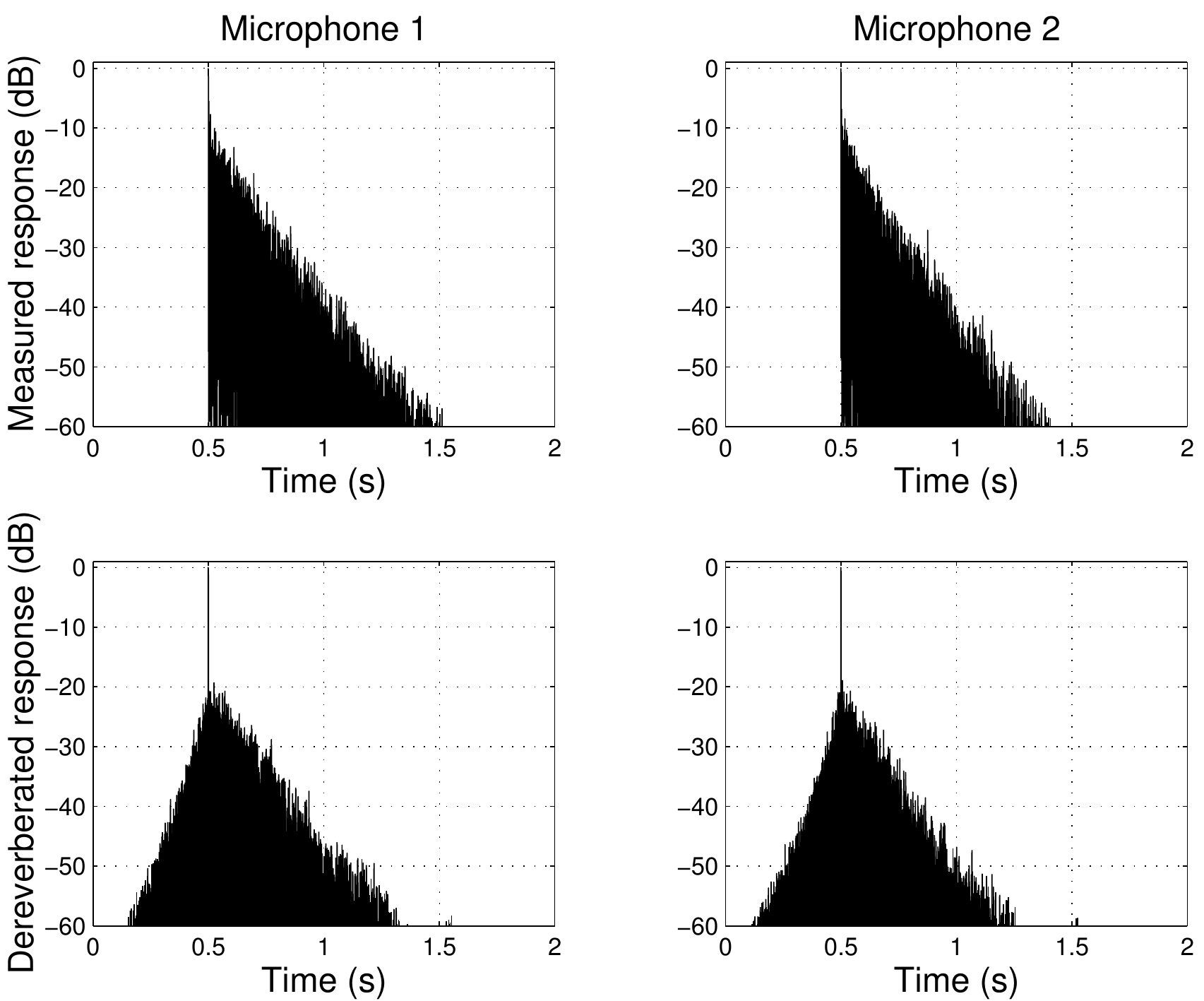}}
\caption{ Decibel plots of measured and dereverberated impulse responses ($2 \times 2$ matrix inversion).  An exponential window with time constant $\tau=0.06$ s is applied to the simulated impulse response $\hat{g}(t)$.}
\label{fig:out_mat_dB}
\end{figure}

\begin{table}
\center
\caption{\label{tab:DR_mat}Dereverberation ratios in dB of
         dereverberated impulse responses; matrix inversion.}
\begin{tabular}{ccc}
\hlinewd{2pt}
             & Microphone 1 & Microphone 2 \\\hline
$DR(\infty)$ & $8.64$       & $12.04$      \\
$DR(100$ ms$)$ & $8.83$       & $12.19$      \\
\hlinewd{2pt}
\end{tabular}
\end{table}

\begin{figure}
\centerline{\includegraphics[width=4.5in]{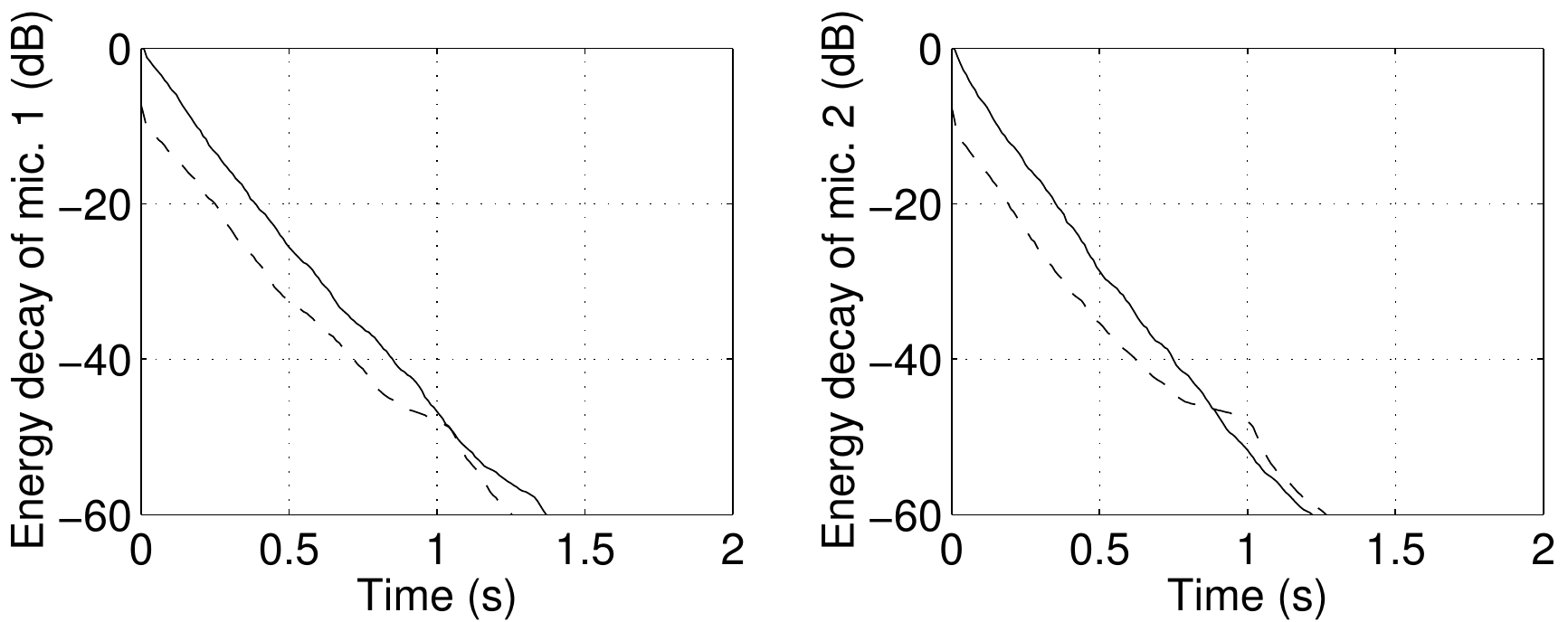}}
\caption{ Integrated energy decay curves for $2 \times 2$ matrix inversion.  Solid line is measured response; dashed line is dereverberated response.}
\label{fig:dec_mat}
\end{figure}

\begin{table}
\center
\caption{\label{tab:rt_mat}Remainder reverberation times $T_{10}$, $T_{20}$, and $T_{60}$, in seconds, of measured and dereverberated impulse responses; matrix inversion.}
\begin{tabular}{ccccc}
\hlinewd{2pt}
         & \multicolumn{2}{c}{Measured responses} & \multicolumn{2}{c}{Dereverberated responses} \\
         & Microphone 1 & Microphone 2 & Microphone 1 & Microphone 2 \\\hline
$T_{10}$ & $0.19$       & $0.16$       & $0.02$       & $0.02$       \\
$T_{20}$ & $0.39$       & $0.36$       & $0.26$       & $0.20$       \\
$T_{60}$ & $1.37$       & $1.22$       & $1.26$       & $1.27$       \\
\hlinewd{2pt}
\end{tabular}
\end{table}

\section{Conclusion}
\label{sec:conclusion}

This paper describes experiments in open-loop room response inversion
for the purpose of headphone-free virtual reality audio display. Room
responses were simulated using the image source method and inverted
using a regularized Fourier transform inversion with a modeling delay
of 500 ms. Scalar room response inversion provided an average of 10.1
dB of short-term dereverberation (early echoes within 100~ms of the
direct sound), and 10.4 dB of long-term dereverberation. Matrix room
response inversion (two inputs, two outputs) provided an average of 10.5
dB short-term and 10.3 dB long-term dereverberation.

\section{Acknowledgments}
This work was supported in part by a grant from the University of
Illinois Research Board, and in part by funding from the University of
Illinois Beckman Institute. We thank Hank Kaczmarski for hardware
assistance. 

\bibliographystyle{jasasty}
\bibliography{thebib}

\begin{thebibliography}{10}

\bibitem{Cruz-Niera1993}
C. Cruz-Neira, D.~J. Sandin, and T.~A. DeFanti, ``Surround-screen
  projection-based virtual reality: the design and implementation of the
  {CAVE},'' Proc. ACM Special Interest Group on Computer Graphics and
  Interactive Techniques ({SIGGRAPH}),135--142 (1993).
\vspace{.125in}
\bibitem{Miyoshi1988}
M. Miyoshi and Y. Kaneda, ``Inverse filtering of room acoustics,'' {IEEE}
  Trans. Acoustics, Speech and Signal Proc. {\bf 36}(2), 145--152 (1988).
\vspace{.125in}
\bibitem{Kirkeby1998}
O. Kirkeby, P.~A. Nelson, H. Hamada, and F. Orduna-Bustamente, ``Fast
  deconvolution of multichannel systems using regularization,'' {IEEE} Trans.
  Speech and Audio Processing {\bf 6}(2), 189--194 (1998).
\vspace{.125in}
\bibitem{Allen1979}
J.~B. Allen and D.~A. Berkley, ``Image method for efficiently simulating
  small-room acoustics,'' J. Acoust. Soc. Am. {\bf 65}(4), 912--915 (1979).
\vspace{.125in}
\bibitem{Neely1979}
S.~T. Neely and J.~B. Allen, ``Invertibility of a room impulse response,'' J.
  Acoust. Soc. Am. {\bf 66}(1), 165--169 (1979).
\vspace{.125in}
\bibitem{Omoto2002}
A. Omoto, C. Hiratsuka, H. Fujita, T. Fukushima, M. Nakahara, and K. Fujiwara,
  ``Similarity evaluation of room acoustic impulse responses: Visual and
  auditory impressions,'' J. Audio. Eng. Soc. {\bf 50}(6), 451--457 (2002).
\vspace{.125in}
\bibitem{ISO-3382}
{ISO 3382}, {\sl Acoustics - Measurement of the reverberation time of rooms
  with reference to other acoustical parameters}, {ISO}, 1997.
\vspace{.125in}
\bibitem{Schroeder1979}
M.~R. Schroeder, ``Integrated-impulse method measuring sound decay without
  using impulses,'' J. Acoust. Soc. Am. {\bf 66}(2), 497--500 (1979).
\vspace{.125in}
\bibitem{Rife1989}
D.~D. Rife and J. Vanderkooy, ``Transfer-function measurement with
  maximum-length sequences,'' J. Audio. Eng. Soc. {\bf 37}(6), 419--444 (1989).
\vspace{.125in}
\bibitem{Vanderkooy1994}
J. Vanderkooy, ``Aspects of {MLS} measuring systems,'' J. Audio. Eng. Soc. {\bf
  42}(4), 219--231 (1994).
\vspace{.125in}
\bibitem{Stan2002}
G.-B. Stan, J.-J. Embrechts, and D. Archambeau, ``Comparison of different
  impulse response measurement techniques,'' J. Audio. Eng. Soc. {\bf 50}(4),
  249--262 (2002).
\vspace{.125in}
\bibitem{Dunn1993}
C. Dunn and M.~O. Hawksford, ``Distortion immunity of {MLS}-derived impulse
  response measurements,'' J. Audio. Eng. Soc. {\bf 41}(5), 314--335 (1993).
\vspace{.125in}
\bibitem{Ream1970}
N. Ream, ``Nonlinear identification using inverse-repeat m sequences,'' Proc.
  IEE (London) {\bf 117}(1), 213--218 (1970).
\vspace{.125in}
\bibitem{Briggs1966}
P.~A.~N. Briggs and K.~R. Godfrey, ``Pseudorandom signals for the dynamic
  analysis of multivariable systems,'' Proc. IEE {\bf 113}, 1259--1267 (1966).
\vspace{.125in}
\bibitem{Simpson1966}
H.~R. Simpson, ``Statistical properties of a class of pseudorandom sequence,''
  Proc. IEE (London) {\bf 113}, 2075--2080 (1966).
\vspace{.125in}
\bibitem{Bleakley1995}
C. Bleakley and R. Scaife, ``New formulas for predicting the accuracy of
  acoustical measurements made in noisy environments using the averaged
  {m}-sequence correlation technique,'' J. Acoust. Soc. Am. {\bf 97}(2),
  1329--1332 (1995).
\vspace{.125in}
\bibitem{Alrutz1983}
H. Alrutz and M.~R. Schroeder, ``A fast {Hadamard} transform method for the
  evaluation of measurements using pseudorandom test signals,'' Proc. 11th Int.
  Congress on Acoustics (Paris) {\bf 6}, 235--238 (1983).
\vspace{.125in}
\bibitem{Cohn1977}
M. Cohn and A. Lempel, ``On fast {M}-sequence transforms,'' {IEEE} Trans.
  Inform. Theory {\bf 23}(1), 135--137 (1977).
\vspace{.125in}
\bibitem{Davies1966a}
W.~D.~T. Davies, ``Generation and properties of maximum-length sequences, part
  1,'' Control {\bf 10}(96), 302--304 (1966).
\vspace{.125in}
\bibitem{Davies1966b}
W.~D.~T. Davies, ``Generation and properties of maximum-length sequences, part
  2,'' Control {\bf 10}(97), 364--365 (1966).
\vspace{.125in}
\bibitem{Davies1966c}
W.~D.~T. Davies, ``Generation and properties of maximum-length sequences, part
  3,'' Control {\bf 10}(98), 431--433 (1966).
\vspace{.125in}
\bibitem{Rife1992}
D.~D. Rife, ``Modulation transfer function measurement with maximum-length
  sequences,'' J. Audio. Eng. Soc. {\bf 40}(10), 779--790 (1992).
\vspace{.125in}
\bibitem{Vorlander1997}
M. Vorl{\"a}nder and M. Kob, ``Practical aspects of {MLS} measurements in
  building acoustics,'' Appl. Acoustics {\bf 52}(3-4), 239--258 (1997).
\vspace{.125in}
\bibitem{Burkard1990}
R. Burkard, Y. Shi, and K.~E. Hecox, ``A comparison of maximum length and
  {Legendre} sequences for the derivation of brain-stem auditory-evoked
  responses at rapid rates of stimulation,'' J. Acoust. Soc. Am. {\bf 87}(4),
  1656--1664 (1990).
\vspace{.125in}
\bibitem{Aoshima1981}
N. Aoshima, ``Computer-generated pulse signal applied for sound measurement,''
  J. Acoust. Soc. Am. {\bf 69}(5), 1484--1488 (1981).
\vspace{.125in}
\bibitem{Berkhout1980}
A.~J. Berkhout, D. de~Vries, and M.~M. Boone, ``A new method to acquire impulse
  responses in concert halls,'' J. Acoust. Soc. Am. {\bf 68}(1), 179--183
  (1980).
\vspace{.125in}
\bibitem{Farina2002}
A. Farina, ``Simultaneous measurement of impulse response and distortion with a
  swept-sine technique,'' J. Audio. Eng. Soc. {\bf 48}, 350 (2000).
\vspace{.125in}
\bibitem{Rabenstein1999}
R. Rabenstein and A. Zayati, ``A direct method to computational acoustics,''
  {IEEE} Proc. Int. Conf. Acoustics, Speech and Signal Processing (ICASSP 99)
  {\bf 2}, 957--960 (1999).
\vspace{.125in}
\bibitem{Schetelig1998}
T. Schetelig and R. Rabenstein, ``Simulation of three-dimensional sound
  propagation with multidimensional wave digital filters,'' {IEEE} Proc. Int.
  Conf. Acoustics, Speech and Signal Processing (ICASSP 98) {\bf 6}, 3537--3540
  (1998).
\vspace{.125in}
\bibitem{Krokstad1968}
A. Krokstad, S. Str{\o}m, and S. S{\o}rsdal, ``Calculating the acoustical room
  response by the use of a ray tracing technique,'' J. Sound Vib. {\bf 8}(1),
  118--125 (1968).
\vspace{.125in}
\bibitem{Borish1984}
J. Borish, ``Extension of the image model to arbitrary polyhedra,'' J. Acoust.
  Soc. Am. {\bf 75}(6), 1827--1836 (1984).
\vspace{.125in}
\bibitem{Lee1988}
H. Lee and B.-H. Lee, ``An efficient algoritm for the image model technique,''
  Applied Acoustics {\bf 24}, 87--115 (1988).
\vspace{.125in}
\bibitem{Nelson1992}
P.~A. Nelson, H. Hamada, and S.~J. Elliott, ``Adaptive inverse filters for
  stereophonic sound reproduction,'' {IEEE} Trans. Signal Processing {\bf
  40}(7), 1621--1632 (1992).
\vspace{.125in}
\bibitem{MilitaryResearch}
{\sl Database for assessing the annoyance of the noise of small arms}, {United
  States Army Environmental Hygiene Agency}, 1983.
\vspace{.125in}
\bibitem{Cramer1993}
O. Cramer, ``The variation of the specific heat ratio and the speed of sound in
  air with temperature, pressure, humidity, and {CO}$_2$ concentration,'' J.
  Acoust. Soc. Am. {\bf 93}(5), 2510--2514 (1993).
\vspace{.125in}
\bibitem{Kinsler2000}
L.~E. Kinsler, A.~R. Frey, A.~B. Coppens, and J.~V. Sanders,  {\sl Fundamentals
  of Acoustics}, 4th ed. (Wiley and Sons, Inc., New York, 2000).
\vspace{.125in}
\bibitem{CRC_Handbook}
 {\sl CRC Handbook of Chemistry and Physics}, 79th ed., edited by D.~R. Lide
  (CRC Press, Boca Raton, 1998).
\vspace{.125in}
\bibitem{Schroeder1965}
M.~R. Schroeder, ``New method of measuring reveberation time,'' J. Acoust. Soc.
  Am. {\bf 37}(3), 409--412 (1965).
\vspace{.125in}
\end{thebibliography}
}
\end{document}